\documentclass[twocolumn, emulateapj]{aastex61}
\usepackage{amsmath, hyperref, textcomp, gensymb, float}

\shorttitle{Infrared Transit Photometry for Four \textit{Kepler} TTV Systems}
\shortauthors{Vissapragada et al.}

\begin{document}

\title{Diffuser-Assisted Infrared Transit Photometry for Four Dynamically Interacting \textit{Kepler} Systems}

\correspondingauthor{Shreyas Vissapragada}
\email{svissapr@caltech.edu}

\author[0000-0003-2527-1475]{Shreyas Vissapragada}

\affil{Division of Geological and Planetary Sciences,
California Institute of Technology, 1200 California Blvd, Pasadena, CA 91125, USA}
\author{Daniel Jontof-Hutter}
\affil{Department of Physics, University of the Pacific, 3601 Pacific Avenue, Stockton, CA 95211, USA}

\author[0000-0002-1836-3120]{Avi Shporer}
\affil{Kavli Institute for Astrophysics and Space Research, Massachusetts Institute of Technology, Cambridge, MA 02139, USA}

\author{Heather A. Knutson}

\affil{Division of Geological and Planetary Sciences, California Institute of Technology, 1200 California Blvd, Pasadena, CA 91125, USA}

\author{Leo Liu}
\affil{Department of Astronomy \& Astrophysics, The Pennsylvania State University, 525 Davey Lab, University Park, PA 16802, USA}
\affil{Center for Exoplanets \& Habitable Worlds, University Park, PA 16802, USA}

\author{Daniel Thorngren}
\affil{Department of Physics, University of California, Santa Cruz, CA, USA}

\author[0000-0002-1228-9820]{Eve J. Lee}
\affil{TAPIR, Walter Burke Institute for Theoretical Physics, Mailcode 350-17, Caltech, Pasadena, CA 91125, USA}
\affil{Department of Physics and McGill Space Institute, McGill University, 3550 rue University, Montreal, QC, H3A 2T8, Canada}

\author[0000-0003-1728-8269]{Yayaati Chachan}
\affil{Division of Geological and Planetary Sciences, California Institute of Technology, 1200 California Blvd, Pasadena, CA 91125, USA}
\author[0000-0002-8895-4735]{Dimitri Mawet}
\affil{Department of Astronomy, California Institute of Technology, 1200 California Blvd, Pasadena, CA 91125, USA}
\affil{Jet Propulsion Laboratory, California Institute of Technology, 4800 Oak Grove Dr, Pasadena, CA 91109, USA}
\author[0000-0001-6205-9233]{Maxwell A. Millar-Blanchaer}
\affil{Jet Propulsion Laboratory, California Institute of Technology, 4800 Oak Grove Dr, Pasadena, CA 91109, USA}
\author[0000-0002-5408-4954]{Ricky Nilsson}
\affil{Department of Astronomy, California Institute of Technology, 1200 California Blvd, Pasadena, CA 91125, USA}
\author[0000-0002-1481-4676]{Samaporn Tinyanont}
\affil{Department of Astronomy, California Institute of Technology, 1200 California Blvd, Pasadena, CA 91125, USA}
\author[0000-0002-1871-6264]{Gautam Vasisht}
\affil{Jet Propulsion Laboratory, California Institute of Technology, 4800 Oak Grove Dr, Pasadena, CA 91109, USA}
\author[0000-0001-6160-5888]{Jason T. Wright}
\affil{Department of Astronomy \& Astrophysics, The Pennsylvania State University, 525 Davey Lab, University Park, PA 16802, USA}
\affil{Center for Exoplanets \& Habitable Worlds, University Park, PA 16802, USA}
\begin{abstract}
We present ground-based infrared transit observations for four dynamically interacting \textit{Kepler} planets, including Kepler-29b, Kepler-36c, KOI-1783.01, and Kepler-177c, obtained using the Wide-field Infrared Camera on the Hale 200" telescope at Palomar Observatory. By utilizing an engineered diffuser and custom guiding software, we mitigate time-correlated telluric and instrumental noise sources in these observations.  We achieve an infrared photometric precision comparable to or better than that of space-based observatories such as the \textit{Spitzer Space Telescope}, and detect transits with greater than 3$\sigma$ significance for all planets. For Kepler-177c ($J=13.9$) our measurement uncertainties are only $1.2\times$ the photon noise limit and 1.9 times better than the predicted photometric precision for \textit{Spitzer} IRAC photometry of this same target. We find that a single transit observation obtained $4-5$ years after the end of the original \textit{Kepler} mission can reduce dynamical mass uncertainties by as much as a factor of three for these systems. Additionally, we combine our new observations of KOI-1783.01 with information from the literature to confirm the planetary nature of this system. We discuss the implications of our new mass and radius constraints in the context of known exoplanets with low incident fluxes, and we note that Kepler-177c may be a more massive analog to the currently known super-puffs given its core mass (3.8$\pm0.9M_\Earth$) and large gas-to-core ratio (2.8$\pm0.7$). Our demonstrated infrared photometric performance opens up new avenues for ground-based observations of transiting exoplanets previously thought to be restricted to space-based investigation.
\end{abstract}

\keywords{techniques: photometric -- planets and satellites: fundamental parameters -- planets and satellites: individual (Kepler-29b, Kepler-36c, KOI-1783.01, Kepler-177c)}

\section{Introduction \label{sec:intro}}
The \textit{Kepler} mission \citep{b10} has revealed thousands of transiting exoplanets and exoplanet candidates over the past decade, many of which reside in multi-planet systems. Dynamical interactions between planets in these systems cause deviations from the expected Keplerian behavior that can change both the timing and duration of transits \citep{a05, hm05, af17}. In systems where planetary periods are close to integer multiples of each other -- in other words, for planets close to or occupying mean motion resonances -- the amplitude of transit timing variations (TTVs) and transit duration variations (TDVs) may become observable and reveal the dynamical architecture of the system. Approximately 10\% of Kepler Objects of Interest (KOIs) exhibit significant long-term TTVs \citep{h16}. Most of these planets are on $\lesssim 100$~day orbits, with eccentricities of a few percent and sizes ranging from 1-10~$R_\Earth$ \citep{h16, hl17}.

TTV analyses have yielded a wealth of information about the properties of \textit{Kepler} multi-planet systems, but arguably their most valuable contribution to date has been estimates of planet masses and densities for systems that are not amenable to characterization using the radial velocity (RV) technique \citep[e.g.][]{wl13, jh16, hl17}. These density constraints are especially critical for interpreting the bimodal radius distribution observed for close-in planets, which peaks at approximately 1.3 and 2.5 $R_\Earth$ \citep{f17b,fp18}. It has been suggested that this distribution is well-matched by models in which a subset of highly irradiated rocky planets have lost their primordial atmospheres while more distant planets retain modest (few percent in mass) hydrogen-rich atmosphere that inflate their observed radii \citep{ow13, lf13, lf14, f17b, ow17, fp18}. Measuring the bulk density of planets in this size regime is thus a direct test of these photoevaporative models.  

\begin{figure}[ht!]
    \centering
    \includegraphics[width = 0.45\textwidth]{{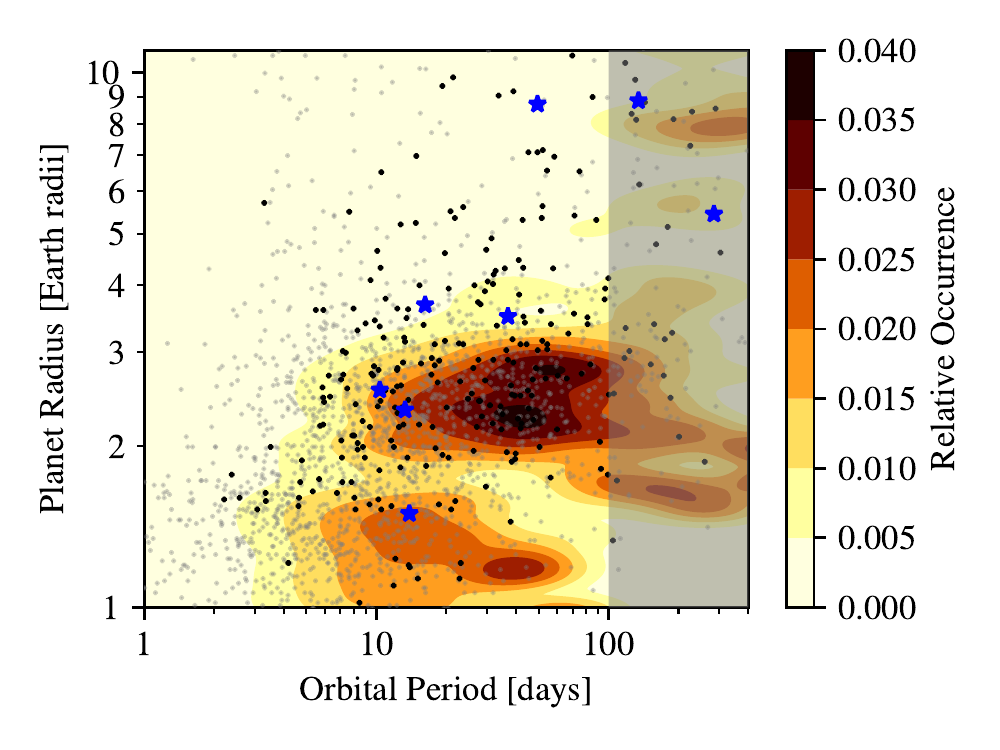}}
    \caption{Planet radius as a function of orbital period for all non-TTV \textit{Kepler} planets (gray points) and the \textit{Kepler} TTV sample (black points), along with the dynamically interacting planets with improved masses from this work (blue stars). The colored contours are the relative planet occurrence contours calculated by \citet{fp18}, and the gray highlighted region denotes the region of low completeness at $P > 100$ days.}
    \label{fultonTTV}
\end{figure}
In Figure~\ref{fultonTTV}, we plot all confirmed \textit{Kepler} planets (with those exhibiting TTVs specially marked) on the radius-period plane, following \citet{f17b}. In general, the TTV sample allows for characterization of planets that are $1.75R_\Earth$ and larger (on the sub-Neptune side of the bimodal radius distribution), with periods longer than a week. While the radial velocity technique is most sensitive to short-period planets with relatively high densities, TTV observations are well-suited to characterizing long period and/or low-density planets, making it an important tool for probing this region of parameter space \citep{s16a, mm17}. Indeed, this technique has already revealed the existence of a separate sub-population of ``super-puffs,'' a rare class of super-Earths with very low bulk densities and relatively long orbital periods \citep{m14, jh14}. Unlike the broader super-Earth population, which some studies argue could have formed in situ, it is thought that these planets may have accreted their envelopes at large stellocentric distances and then migrated inward to their current locations in resonant chains \citep{ih12, l14, g16, lc16, s18b}. 

These previous studies showcase the crucial role of \textit{Kepler} TTVs in testing theories of planet formation and evolution. The failure of \textit{Kepler}'s second reaction wheel in 2013, however, effectively limited the baseline of these TTV analyses to four years. This makes it particularly challenging to constrain masses and bulk densities for long-period planets with a relatively small set of measured transits during this four-year period. In addition, uncertainties in the orbital solutions grow over time, making future in-transit observations (for instance, those aimed at atmospheric characterization) increasingly difficult to schedule with confidence.

These problems can be ameliorated with ground- or space-based follow-up observations \citep{p18, w18a}. However, many of the \textit{Kepler} planets exhibiting TTVs orbit faint ($V > 12$) stars, making it difficult to achieve the required photometric precision using existing space-based facilities with small apertures, such as the \textit{Spitzer Space Telescope}. Additionally, \textit{Spitzer} will be decommissioned in January 2020, necessitating an alternative approach to follow-up observations. Although ongoing observations by the \textit{Transiting Exoplanet Survey  Satellite} \citep[\textit{TESS};][]{r15} are expected to recover a few hundred \textit{Kepler} planets \citep{c18}, short-cadence data from the nominal mission will only improve the mass uncertainties for 6-14 of the $\sim$150 currently known \textit{Kepler} TTV planets \citep{g19}. This is due to the limited photometric precision and relatively short baseline of \textit{TESS} relative to \textit{Kepler}. While \textit{TESS} is expected to recover additional transits in an extended mission scenario, these detections will still constitute less than 20\% of the overall \textit{Kepler} TTV sample \citep{g19}.

Ground-based observatories can in principle recover transits for faint \textit{Kepler} stars with long period planets, and coordinated multi-observatory campaigns have shown promise in achieving the requisite phase coverage \citep{f18, v18b, w18a}. However, their photometric precisions are typically limited by low observing efficiencies and the presence of time-correlated noise due to imperfect guiding and point-spread function (PSF) variations \citep{z14, c15, s17}.  These difficulties can be mitigated by using diffusers to control the shape of the point spread function (PSF) and spread out light from the star over a larger area. Diffusers have already been installed on several ground-based telescopes and have been shown to achieve significantly better photometric precision than more traditional observing techniques \citep{s17, s18, v19}.

Here, we present diffuser-assisted TTV follow-up observations of four \textit{Kepler} planets in dynamically interacting systems. We discuss our sample selection methodology and our observations of the four-planet sample with the Wide-field InfraRed Camera \citep[WIRC;][]{w03} in Section \ref{sec:obs}. In Section \ref{sec:methods}, we describe our image calibration, data reduction, light curve modeling, and dynamical modeling methods. We then present our results for each system in Section \ref{sec:results}, along with some brief comments on the general performance of our instrument. In Section \ref{sec:discussion}, we discuss some of the scientific implications of our new dynamical mass constraints within the broader exoplanet population, and we conclude with a summary of our results and a look towards future possibilities in Section \ref{sec:conc}.

\section{Observations \label{sec:obs}}
\subsection{Sample Selection\label{sec:sample}}
In this study we focused on the set of multi-planet systems from the original \textit{Kepler} survey. We began by estimating the expected TTV signal strength for all planet pairs  in order to identify the systems most likely to exhibit strong transit timing variations. We estimated the minimum mass of a planet from its radius, and then estimated the chopping signal and near-first order resonant TTV signal for planet pairs given their orbital periods. We then use the number of transits and the transit timing uncertainty to estimate a minimum TTV signal-to-noise ratio (SNR) in the limit of circular orbits. For systems exhibiting TTVs with high SNRs, we performed dynamical fits to the long cadence transit times in \citet{rt15}. We fit five parameters per planet, including the orbital period and phase at a chosen epoch, the two eccentricity vector components, and the dynamical mass. We then mapped the resulting posterior using Differential Evolution Markov Chain Monte Carlo sampling \citep{jh16}. Since mutual inclinations are a second-order effect for the TTV amplitude, we assumed coplanarity in our models \citep{l12c, nv14, jh16}. We then forward modeled sample solutions for each system in order to identify those with the most strongly diverging TTV predictions. A detailed report of our forward modeling is in preparation. 

We selected targets for our WIRC program from the subset of systems with strongly detected TTVs and dynamical solutions that diverged measurably in the years following the end of the primary \textit{Kepler} mission. We excluded systems where the $1\sigma$ range of predicted transit times at the epoch of our proposed WIRC observation was greater than one hour, as this meant that there was a significant possibility that the transit might occur outside our window of observability. In order to ensure that the measured transit time was likely to provide a useful constraint on the dynamical fit we also calculated the expected timing precision of a new WIRC observation and excluded systems where this uncertainty was greater than the $1\sigma$ range in predicted transit times.

Within this sample of systems, we searched for targets with an ingress and/or egress visible from Palomar between August 2017 and May 2018. We then ranked the targets in our sample based on predicted signal-to-noise ratio (SNR) scaled from early WIRC commissioning data \citep{s17}, and prioritized observations of the highest SNR targets. We ultimately obtained high-quality light curves for four confirmed and candidate planets from this ranked list, including: Kepler-29b, Kepler-36c, KOI-1783.01, and Kepler-177c. The predicted mid-transit times for these planets are shown in Table \ref{table1}.

\begin{deluxetable*}{ccccccccc}[t!]
\tabletypesize{\scriptsize}
\tablecaption{Observational parameters for our four nights of data collection. \label{table1}}
\tablehead{\colhead{Star} & \colhead{\textit{J} mag\tablenotemark{a}} & Date  & \colhead{Start Time} & \colhead{End Time} & \colhead{Event Time\tablenotemark{b}} & \colhead{Event Duration} & \colhead{Start/Min/End Airmass} &\colhead{Exposure Time} \\ & & (UTC) & (UTC) & (UTC) & (UTC) & (hr) & & (s)}
\startdata
Kepler-29 & 14.13 & 2017 August 25 & 05:35:24 & 11:57:00 & 08:26:53 & 3.046 & 1.03/1.03/3.01 & 25\phm{\tablenotemark{d}} \\
Kepler-36 & 11.12 & 2017 September 27 & 03:06:20 & 08:55:42 & 09:52:34 & 7.461 & 1.04/1.04/2.50& 16\tablenotemark{d} \\
KOI-1783 & 12.92 & 2018 April 21  & 08:19:42 & 12:04:05& 07:07:51 & 5.871 & 1.73/1.05/1.05 & 20\phm{\tablenotemark{d}} \\
Kepler-177 & 13.86 & 2018 May 4 & 07:17:36 & 12:09:04 & 10:30:49 & 5.245 & 1.73/1.02/1.02 & 75\tablenotemark{e}\\
\enddata
\tablenotetext{a}{\textit{J} band magnitudes from the 2MASS catalogue \citep{c03}.}
\tablenotetext{b}{Predicted mid-transit time.}
\tablenotetext{d}{4 co-adds of 4 second exposures.}
\tablenotetext{e}{3 co-adds of 25 second exposures.}
\end{deluxetable*}

\subsection{New WIRC Observations\label{sec:newobs}}
We observed our four selected systems in \textit{J} band with WIRC, which is located at the prime focus of the Hale 200" telescope at Palomar Observatory \citep{w03}. The current 2048 $\times$ 2048 pixel Hawaii-II HgCdTe detector was installed in January 2017, along with 32-channel readout electronics that allow for a read time of 0.92 s \citep{t19}. The instrument has an 8\farcm7 $\times$ 8\farcm7 field of view with a pixel scale of 0\farcs2487, ensuring that (at least for the magnitude range in our sample) there are always on the order of ten stars with comparable brightness contained within the same field of view as our target star. 

We utilize the custom near-infrared Engineered Diffuser described in \citet{s17} to mitigate time-correlated noise from PSF variations and improve our observing efficiency. The diffuser delivers a top-hat PSF with a full width at half maximum (FWHM) of 3\arcsec. We also minimize the time-correlated noise contribution from flat-fielding errors by utilizing precision guiding software \citep{z14}. WIRC does not have a separate guide camera, but instead guides on science images by fitting 2D Gaussian profiles to comparison stars and determining guiding offsets on each image. For these observations, we find that the position of the star typically varies by less than 2-3 pixels over the course of the night, with the largest position drift occurring at high airmass where accurate centroid measurements become more challenging.

Dates, times, and airmasses for each observation are reported in Table \ref{table1}. For Kepler-29, Kepler-36, and Kepler-177, we observed continuously during the observation windows. During our observation of KOI-1783 there were three breaks in data acquisition due to a malfunctioning torque motor causing a temporary loss of telescope pointing.

Exposure times are also reported in Table \ref{table1}, and were chosen to keep the detector in the linear regime. WIRC commissioning tests have shown the detector to be linear to $\sim0.5\%$ at 22,000 ADU \citep{t19}. When choosing exposure times, we aimed to keep the maximum count level at or below 20,000 ADU in order to accommodate potential changes in airmass and sky background. In some cases, frames were co-added during the night to increase observing efficiency as noted in Table \ref{table1}.

\section{Data Reduction and Model Fits \label{sec:methods}}

\subsection{Image Calibration and Photometry \label{sec:calibration}}
For each night, we construct a median dark frame and a flat field. During the construction of the dark and flat, we also construct a global bad pixel map with the procedure described by \citet{t19}. Each image is dark subtracted and flat-fielded, and each bad pixel is replaced with the median of the 5 pixel $\times$ 5 pixel box surrounding the errant value. The total number of bad pixels is approximately 0.6\% of the full array \citep{t19}. During the calibration sequence, mid-exposure times are converted to Barycentric Julian Date in the Barycentric Dynamical Time standard (BJD$_\mathrm{TDB}$), following the recommendation of \citet{e10}. All of the above steps are performed by the WIRC Data Reduction Pipeline, which was originally developed to automatically handle large sets of polarimetric data \citep{t19}.

We perform aperture photometry using the \texttt{photutils} package \citep{b16b}. We begin by using the first science image as a ``finding frame'' and detect sources using the \texttt{DAOStarFinder} function \citep[based on][]{s87}. Sources that are close to the detector edge and those with overlapping apertures are removed automatically. The target star is registered by comparison to an Aladin Lite finding chart \citep{b00, b14}. We then perform the photometry using a range of circular apertures with radii ranging between 6 and 18 pixels in one pixel steps, using the same aperture for all stars in each image. With WIRC's $\sim0\farcs25$/pixel scale, the diffuser is expected to deliver stellar PSFs with a FWHM of 12 pixels, but the actual FWHM changes with stellar brightness. For each image, we calculate and subtract the median background via iterative 3$\sigma$ clipping with the \texttt{sigma\_clipped\_stats} function in \texttt{astropy} with a five-iteration maximum specified \citep{a13, a18}. After this, we re-calculate the source centroids via iterative flux-weighted centroiding and shift apertures accordingly for each individual image. The local sky background is then estimated using an annular region around each source with inner radius of 20 pixels and outer radius of 50 pixels. We find that iterative sigma-clipping of this background region (this time with a $2\sigma$ threshold) is sufficient to reconstruct the mean local background, even though the fields are fairly crowded.

After raw light curves are obtained for each aperture size, we choose the ten comparison stars that best track the time-varying flux of the target star (i.e. those that have the minimal variance from the target star). We clean the target and comparison light curves by applying a moving median filter (of width 10 data points) to the target star dataset and removing 3$\sigma$ outliers. We then select the optimal aperture by minimizing the root mean square (RMS) scatter after the light curve fitting described in the next section. Our optimal aperture radii were 8 pixels for Kepler-29b, 14 pixels for Kepler-36c, 10 pixels for KOI-1783.01, and 10 pixels for Kepler-177c. We find that our preferred apertures for each target increase in size with increasing stellar brightness, and all preferred apertures are comparable in size to the aforementioned 12 pixel FWHM expected for the diffuser.

\subsection{\textit{Kepler} Light Curves}
\label{sec:kepler}
Of the four planets in our sample, only one (Kepler-29b) had a transit duration short enough to allow us to observe a full transit; for the other three planets our observations spanned ingress or egress, but not both. This introduces a degeneracy between the mid-transit time and transit duration (parameterized here by the inclination and semi-major axis) in our fits to these four transits. We resolve this degeneracy by carrying out joint fits with the original \textit{Kepler} photometry, where we assume common values for the transit depth $(R_\mathrm{p}/R_\star)^2$, the inclination $i$, and the scaled semi-major axis $a/R_\star$. Although we would expect the transit depth to vary as a function of wavelength if any of these planets have atmospheres, the maximum predicted magnitude for this variation (corresponding to a cloud-free, hydrogen-rich atmosphere) is much smaller than our expected measurement uncertainty for the change in transit depth $(R_\mathrm{p}/R_\star)^2$ between the optical \textit{Kepler} band and our $J$ band photometry. This effect would be strongest for the low-density planet Kepler-177c, but even then, the maximal variation is of order 200 ppm versus our WIRC $J$ band precision of roughly 1300 ppm. We found that constraining the transit depth to the \textit{Kepler} value resulted in smaller transit timing uncertainties for our partial transit observations, which otherwise exhibited correlations between the transit depth, the transit time, and the linear trend in time. 

 We processed the \textit{Kepler} long-cadence simple aperture photometry (SAP) light curves for each star in our sample using the \texttt{kepcotrend} function in the \texttt{PyKE} package \citep{sb12}. To avoid errors in light curve shape introduced by assuming a linear ephemeris, we cut out individual light curves from the cotrended \textit{Kepler} data using lists of individual transit times from \citet{h16} when possible and otherwise using \cite{rt15}. We selected our trim window to provide two transit durations of both pre-ingress and post-egress baseline. After dividing out a linear trend fit to the out-of-transit baseline for each light curve, we combined all transits into a single transit light curve with flux as a function of time from transit center. 
 
 This process assumes that TDVs do not strongly bias our retrieved transit shapes. For systems with large amplitude TDVs it may become necessary to perform photodynamical modeling in order to properly treat the time-varying transit shape \citep[e.g.][]{f18}. However, \cite{h16} examined data spanning the full length of the \textit{Kepler} mission and did not detect TDVs for any of the targets in our sample. To further justify our assumption that TDVs have a negligible impact on the measured signals, we calculated the expected TDV amplitude for Kepler-177c (a planet with long period and large impact parameter that is more prone to nodal precession). The maximum TDV amplitude is of order 0.1 hr over the 10 year baseline. The WIRC data alone are not sensitive to transit duration changes on this timescale, since we only detect ingress or egress for most transits. Additionally, the precision on the transit timing in the joint fits tend to be much more uncertain than 0.1 hr, meaning that TDV effects will not compromise our final TTV constraints.  We conclude that we can safely ignore TDVs in our treatment of these data.

\subsection{Light Curve Fitting \label{sec:wirckepmodeling}}
To fit the \textit{Kepler} and WIRC light curves, we first constructed light curve models defined by observed quantities and fit parameters. We then constructed appropriate likelihood and prior functions and sampled the resultant posterior probability numerically to obtain estimates of the best-fit parameters and their associated uncertainties. The outputs of the WIRC photometry pipeline are an array of times $\vec{t} = (t_1, t_2,...,t_n)$, the target data array $\vec{y} = (y_1, y_2, ..., y_n)$ (with $y_i$ referring to the measurement at time $t_i$), and comparison star arrays $\vec{x}_j = (x_1, x_2, ..., x_n)$. Collectively, the comparison stars define a matrix $\mathbf{X}$, with one comparison star $\vec{x}_j$ in each row of the matrix.

We aim to fit the target $\vec{y}$ with a model $\vec{M}$ that depends on the depth of the transit $(R_\mathrm{p}/R_\star)^2$, the transit center time $t_0$, the inclination $i$, the ratio of semi-major axis to stellar radius $a/R_\star$, and a linear trend in time $\alpha$. That model can be written as follows \citep[loosely following the notation of][]{dl18}:

\begin{equation}
    \vec{M} = [\alpha \vec{t} + \vec{S}]\times\vec{T}_\mathrm{WIRC}((R_\mathrm{p}/R_\star)^2, t_0, i, a/R_\star),
    \label{model}
\end{equation}
where $\vec{S}$ is the systematics model, $\vec{T}_\mathrm{WIRC}$ is the transit model, and the multiplication is meant to denote a pointwise product. We use the \texttt{batman} code to construct the transit model \citep{k15b} and fix the planet eccentricities to zero. The eccentricities of multi-planet \textit{Kepler} systems are typically small, with a population mean of $\bar{e}=0.04^{+0.03}_{-0.04}$ \citep{x16}, and the effect of these eccentricities on the shape of the transit light curve is negligible for these data. We use four-parameter nonlinear limb darkening coefficients from \citet{cb11}, assuming stellar parameter values from \citet{p17} that are reproduced in Table \ref{stellar}.

\begin{deluxetable*}{cccccc}[bht!]
\tabletypesize{\scriptsize}
\tablecaption{Stellar parameters for the stars in our sample. \label{stellar}}
\tablehead{\colhead{Target} & \colhead{$T_\mathrm{eff}$} & \colhead{[Fe/H]} & \colhead{$\log(g)$} & \colhead{$M_\star$} & \colhead{$R_\star$} \\ & (K) & (dex) & (log(cm/s$^2$)) & ($M_\Sun$) & ($R_\Sun$)}
\startdata
Kepler-29 & $5378^{+60}_{-60}$ & $-0.44^{+0.04}_{-0.04}$ & $4.6^{+0.1}_{-0.1}$ & $0.761^{+0.024}_{-0.028}$ & $0.732^{+0.033}_{-0.031}$ \\
Kepler-36 & $5979^{+60}_{-60}$ & $-0.18^{+0.04}_{-0.04}$ & $4.1^{+0.1}_{-0.1}$ & $1.034^{+0.022}_{-0.022}$ & $1.634^{+0.042}_{-0.040}$ \\
 KOI-1783 & $5922^{+60}_{-60}$ & $\phm{-}0.11^{+0.04}_{-0.04}$ & $4.3^{+0.1}_{-0.1}$ & $1.076^{+0.036}_{-0.032}$ & $1.143^{+0.031}_{-0.030}$\\
 Kepler-177 & $5732^{+60}_{-60}$ & $-0.11^{+0.04}_{-0.04}$ & $4.1^{+0.1}_{-0.1}$ & $0.921^{+0.025}_{-0.023}$ & $1.324^{+0.053}_{-0.051}$\\
 \enddata
 \tablecomments{Spectroscopic parameters ($T_\mathrm{eff}$, [Fe/H], and log($g$)) are taken from \citet{f17b}, and physical parameters ($M_\star$ and R$_\star$) are from \citet{fp18}.}
\end{deluxetable*}

\begin{figure}[ht!]
    \centering
    \includegraphics[width = 0.45\textwidth]{{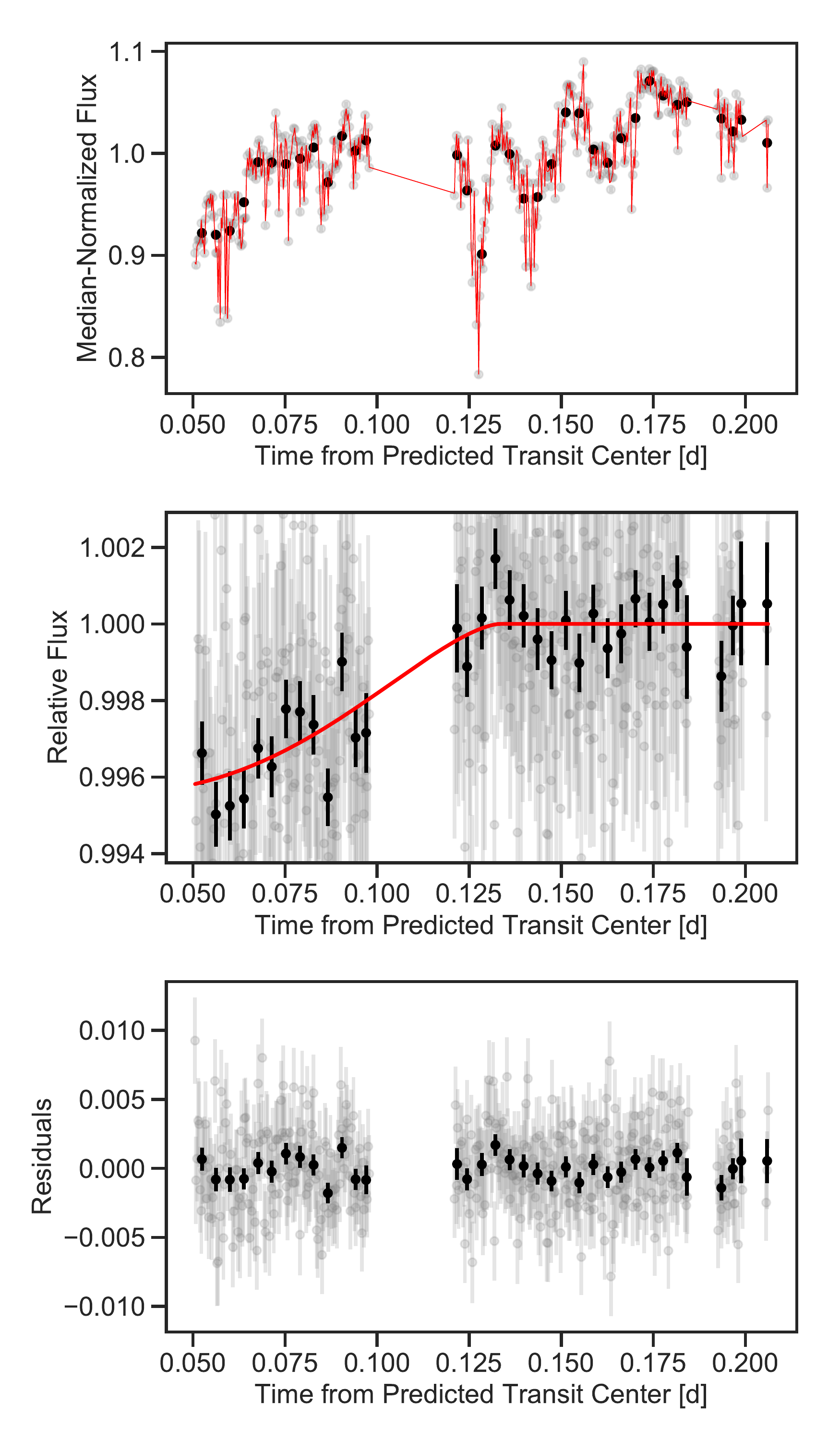}}
    \caption{(Top) Median-normalized photometry for KOI-1783.01, with unbinned data in gray and data binned by a factor of 10 in black. The breaks in data acquisition were due to a malfunctioning torque motor. The best-fit systematic noise model is shown as a red curve. (Middle) Detrended photometry of KOI-1783.01, with the best-fit light curve model now shown in red. (Bottom) Residuals from the light curve fitting of the detrended photometry.}
    \label{tripleplot}
\end{figure}

For ground-based observations, we expect the measured flux from each star to vary as a function of the airmass, centroid drift, seeing changes, transparency variations, and other relevant parameters. However, all of the stars on our wide-field detector should respond similarly to changes in the observing conditions.  In particular, we expect that stars of approximately the same $J$ magnitude and color will track closely with the light curve of our target star. We therefore define our systematics model as a linear combination of comparison star light curves. This allows us to empirically model these effects without explicitly relating them to the relevant atmospheric and telescope state parameters via a parametric model.  We determine the coefficients for the linear combination via a linear regression fit to the target light curve after dividing out the transit light curve model (which we call the ``target systematics'' $\vec{S}_\mathrm{target}$). We calculate new linear coefficients every time the transit light curve is modified. Mathematically, the target systematics can be written:

\begin{equation}
    \vec{S}_\mathrm{target} = \frac{\vec{y}}{\vec{T}_\mathrm{WIRC}((R_\mathrm{p}/R_\star)^2, t_0, i, a/R_\star)} - \alpha \vec{t},
\end{equation}
where division is meant to be pointwise, and the linear regression defining the systematics model can be written:

\begin{equation}
    \vec{S} = \mathbf{P}\vec{S}_\mathrm{target},
\end{equation}
where the projection matrix $\mathbf{P}$ comes from the comparison stars and can be written:

\begin{equation}
    \mathbf{P} = \mathbf{X}^T(\mathbf{X}\mathbf{X}^T)^{-1}\mathbf{X}
    \label{projection}
\end{equation}
Equations (\ref{model})--(\ref{projection}) thus define the model $\vec{M}$ solely as a function of the observed quantities \{$\vec{t},\vec{y},\mathbf{X}$\} and the fit parameters \{$(R_\mathrm{p}/R_\star)^2, t_0, \alpha, i, a/R_\star$\}. To give a sense for how our systematics removal looks in practice, in Figure~\ref{tripleplot} we show the raw and detrended light curves for KOI-1783.01 along with the best systematics and transit models.

As discussed in \S\ref{sec:kepler}, we fit the WIRC photometry jointly with the \textit{Kepler} photometry in order to avoid a strong degeneracy between mid-transit time and transit duration. The \textit{Kepler} photometry consists of an array of times $\vec{t}_{Kep} = (t_1, t_2,...,t_n)$ and the corresponding detrended target data array $\vec{y}_{Kep} = (y_1, y_2, ..., y_n)$. Because these data are already detrended and phased together, the model $\vec{M}_{Kep}$ for the \textit{Kepler} data is simply a \texttt{batman} transit model:
 
\begin{equation}
    \vec{M}_{Kep} = \vec{T}_{Kep}((R_\mathrm{p}/R_\star)^2, i, a/R_\star)
\end{equation}
We supersampled the \textit{Kepler} light curves to 1 min cadence, and used four-parameter nonlinear limb darkening coefficients from \citet{s10} calculated specifically for the \textit{Kepler} bandpass.

Having defined our models, we can now define our likelihood function. We assume measurements to be Gaussian-distributed and uncorrelated (correlated noise is considered briefly in \S\ref{sec:performance}) such that the likelihood takes the form:

\begin{align}
    \log(\mathcal{L}) = &-\frac{1}{2}\sum_i\log(2\pi\sigma_i^2) - \frac{1}{2}\sum_i\Big(\frac{y_i - M_i}{\sigma_i}\Big)^2 \nonumber \\
   &-\frac{1}{2}\sum_i\log(2\pi\sigma_{Kep, i}^2) \nonumber \\
   &- \frac{1}{2}\sum_i\Big(\frac{y_{Kep, i} - M_{Kep, i}}{\sigma_{Kep, i}}\Big)^2 ,
\end{align}
where the uncertainties $\sigma_i$ and $\sigma_{Kep, i}$ are quadrature sums of the Poisson noise from the target star and extra noise terms that can be fitted:

\begin{align}
    \vec{\sigma} &= \sqrt{\vec{\sigma}_\mathrm{phot, WIRC}^2 + \sigma_\mathrm{extra, WIRC}^2}\\
    \vec{\sigma}_{Kep} &= \sqrt{\vec{\sigma}_{\mathrm{phot,} Kep}^2 + \sigma_{\mathrm{extra,} Kep}^2}.
\end{align}
Because the extra noise terms are always positive, we fit for $\log(\sigma_\mathrm{extra, WIRC})$ and $\log(\sigma_{\mathrm{extra,} Kep})$ as a numerical convenience. Also, rather than fitting for $t_0$ itself, we define all times relative to the predicted transit times in Table \ref{table1}, and fit for the offset from that time $\Delta t_0$.

We impose priors on all parameters. They are either Gaussian, taking the functional form:

\begin{equation}
    \log(\mathcal{P}_k) = -\frac{1}{2}\log(2\pi\sigma_k^2) - \frac{1}{2}\Big(\frac{k - \mu_k}{\sigma_k}\Big)^2,
    \label{gaussianprior}
\end{equation}
or uniform, taking the functional form:

\begin{align}
    \log(\mathcal{P}_k) = &\log\Big(\frac{1}{k_\mathrm{max} - k_\mathrm{min}}\Big), \quad k_\mathrm{min} < k < k_\mathrm{max};  \label{uniformprior} \\ 
    &-\infty\ \mathrm{otherwise}\nonumber.
\end{align}
We placed physically motivated Gaussian priors on $a/R_\star$ calculated from the stellar parameters reported by \citet{fp18}, and used uniform priors for all other variables.
We list our priors for the physical fit parameters in Table~\ref{fitting}. 

With the likelihood and priors defined, we can finally write the posterior probability with Bayes' Theorem (up to a constant proportional to the evidence):
\begin{equation}
    \log(\mathrm{Prob}) = \log(\mathcal{L}) + \sum_k\log(\mathcal{P}_k)
    \label{posterior}
\end{equation}
Then, we seek a solution for the fit parameters $(R_\mathrm{p}/R_\star)^2$, $\Delta t_0$, $i$, $a/R_\star$, $\alpha$, $\log(\sigma_\mathrm{extra, WIRC})$, and $\log(\sigma_{\mathrm{extra}, Kep})$ that maximizes $\log(\mathrm{Prob})$. We carry out an initial fit using \texttt{scipy}'s Powell minimizer \citep{j01} and use this solution as a starting point for the affine-invariant ensemble Markov chain Monte Carlo sampler \texttt{emcee} \citep{fm13}. We burn the chains in for $2\times10^3$ steps and then run for 10$^5$ steps. This corresponds to at least 500 integrated autocorrelation times for each parameter. The maximum \textit{a posteriori} parameter estimates with associated 68\% confidence intervals for all model parameters aside from $\alpha$, $\log(\sigma_\mathrm{extra, WIRC})$, and $\log(\sigma_{\mathrm{extra}, Kep})$ are given in Table~\ref{fitting}. The best-fit light curves are shown in Appendix~\ref{ap:lightcurve}. Additionally, we plot the posterior distributions for these parameters in Appendix~\ref{ap:posteriors}. 

\begin{deluxetable*}{ccccccc}
\tabletypesize{\scriptsize}
\tablecaption{Photometric quality statistics for the observations presented in this work. \label{photqual}}
\tablehead{\colhead{Planet} & \colhead{WIRC Transit Coverage} & \colhead{\textit{Kepler} RMS} & \colhead{WIRC RMS} & \colhead{WIRC RMS} & \colhead{WIRC Binned RMS} & \colhead{$\log(\sigma_\mathrm{extra,WIRC})$} \\ & (\%) & (ppm) & (ppm) & ($\times$ photon noise) & ($\times$ photon noise) & }
\startdata
Kepler-29b & 100 & 504 & 4222 & 1.20 & 1.27 & -2.627 \\
Kepler-36c & 41.8 & 75 & 1305 & 2.10 & 2.46 & -2.943 \\
KOI-1783.01 & 33.7 & 157 & 2862 & 1.48 & 1.29 & -2.680 \\
Kepler-177c & 66.9 & 320 & 2403 & 1.22 & 1.46 & -2.851 \\
 \enddata
 \tablecomments{For the binned RMS values, data are binned to 10 minute cadence. Additionally, the \citet{cw09} $\beta$ factor quantifying correlated noise is the binned RMS divided by the unbinned RMS in this parameterization, since both are provided in terms of the photon noise.}
\end{deluxetable*}

\begin{deluxetable*}{cCcccccc}[ph!]
\tablecolumns{8} 
\tabletypesize{\scriptsize}
\tablecaption{System parameters for the joint photometric fits. \label{fitting}}
\tablehead{\colhead{Parameter} & \colhead{Symbol} & \multicolumn{4}{c}{Values} & \colhead{Units} & \colhead{Source} \\
& & \colhead{Kepler-29b} & \colhead{Kepler-36c} & \colhead{KOI-1783.01} & \colhead{Kepler-177c} & &}
\startdata
\cutinhead{Fixed Parameters}
Orbital period & P & 10.3392924 & 16.23192004 & 134.4786723 & 49.41117582 & d & (1, 2)\\
Predicted transit time & t_0 & 2457990.852 & 2458023.9115 & 2458229.7971125 & 2458242.93807 &  BJD & --- \\
Eccentricity & e & 0. & 0. & 0. & 0. & --- & --- \\
\textit{Kepler} limb darkening coefficients & a_1 & \phm{-}0.4959 & \phm{-}0.4639 & \phm{-}0.6034 & \phm{-}0.5716 & --- & (3) \\
 & a_2 & \phm{-}0.0222 & \phm{-}0.3045 & -0.1382& -0.1145 & --- & (3) \\
 & a_3 & \phm{-}0.5708 & \phm{-}0.0751 & \phm{-}0.6330 & \phm{-}0.6579 & 
 --- & (3) \\
 & a_4 & -0.3485 & -0.1251  & -0.3506 & -0.3667 & --- & (3) \\
WIRC limb darkening coefficients & b_1 & \phm{-}0.3634 & \phm{-}0.3982 &\phm{-}0.4832 & \phm{-}0.4421 & --- & (4) \\
 & b_2 & \phm{-}0.5846 & \phm{-}0.5452 & \phm{-}0.2998 & \phm{-}0.3993 &  --- & (4) \\
 & b_3 & -0.6152 & -0.6817 & -0.3634 & -0.4523 &--- & (4) \\
 & b_4 & \phm{-}0.1997 & \phm{-}0.2508 & \phm{-}0.1152 & \phm{-}0.1474 & --- & (4) \\
\cutinhead{Fit Priors}
Transit depth prior & \mathcal{P}_{(R_\mathrm{p}/R_\star)^2} & $\mathcal{U}(0, 2000)$ & $\mathcal{U}(0, 1000)$& $\mathcal{U}(0, 10000)$ & $\mathcal{U}(0, 8000)$ & ppm & --- \\
Transit timing offset prior & \mathcal{P}_{\Delta t_0} & $\mathcal{U}(-100, 100)$ & $\mathcal{U}(-100, 100)$ & $\mathcal{U}(-100, 100)$ & $\mathcal{U}(-100, 100)$ & min & ---\\
Inclination prior & \mathcal{P}_{i} & $\mathcal{U}(85, 90)$ & $\mathcal{U}(85, 90)$ & $\mathcal{U}(85, 90)$ &  $\mathcal{U}(85, 90)$ & \degree & ---\\ 
Scaled semi-major axis prior & \mathcal{P}_{a/R_\star} & $\mathcal{N}(24.906, 1.125)$ & $\mathcal{N}(16.696, 0.436)$ & $\mathcal{N}(99.030, 2.840)$ & $\mathcal{N}(41.649, 1.674)$ & --- & (5)\\
\cutinhead{Fit Posteriors}
Transit depth & (R_\mathrm{p}/R_\star)^2 & 1020$^{+31}_{-34}$ & 425.3$^{+3.8}_{-3.5}$ & 5044$^{+87}_{-64}$ &  3643$^{+55}_{-57}$ & ppm & ---\\
Transit timing offset & \Delta t_0 & -14.3$^{+16.7}_{-2.7}$ & -17.9$^{+11.8}_{-4.7}$
 & 16$^{+10}_{-11}$ &   45.2$^{+8.7}_{-7.1}$ & min & --- \\
Inclination & i &  89.13$^{+0.45}_{-0.23}$ & 89.36$^{+0.45}_{-0.29}$ & 89.4413$^{+0.0076}_{-0.0082}$
 &  88.795$^{+0.037}_{-0.035}$ &  \degree & --- \\
Scaled semi-major axis & a/R_\star & 24.95$^{+1.34}_{-0.91}$ & 16.69$^{+0.26}_{-0.31}$ & 94.8$^{+1.1}_{-1.1}$ & 42.08$^{+1.04}_{-0.94}$ & --- & --- \\
\cutinhead{Derived Parameters}
Planet-star radius ratio & R_\mathrm{p}/R_\star & 0.03194$^{+0.00048}_{-0.00054}$ & 0.02062$^{+0.00009}_{-0.00009}$ & 0.07102$^{+0.00061}_{-0.00045}$ &  0.06036$^{+0.00045}_{-0.00047}$ & -- & -- \\
Impact Parameter & b & 0.379$^{+0.083}_{-0.185}$ & 0.186$^{+0.080}_{-0.131}$ & 0.9239$^{+0.0026}_{-0.0023}$ &  0.8848$^{+0.0056}_{-0.0065}$ & -- & -- \\
Transit duration & T_{14} & 3.041$^{+0.045}_{-0.052}$ &  7.46$^{+0.021}_{-0.017}$ & 5.874$^{+0.039}_{-0.040}$ & 5.243$^{+0.054}_{-0.054}$ & hr & --\\
\enddata
\tablecomments{(1) \citet{m16}, (2) \citet{t18}, (3) \citet{s10}, (4) \citet{cb11}, (5) \citet{fp18}. Also, $\mathcal{N}(a, b)$ indicates a normal (Gaussian) prior with mean $a$ and standard deviation $b$ described by Equation (\ref{gaussianprior}), whereas $\mathcal{U}(a, b)$ indicates a uniform prior with lower bound $a$ and upper bound $b$ described by Equation (\ref{uniformprior}). }
\end{deluxetable*}

\subsection{Dynamical Modeling \label{sec:dynamical}}
Our fits to the ground-based WIRC photometry typically resulted in a non-Gaussian posterior for the mid-transit time. We accounted for these skewed distributions in our dynamical fits by dividing the posteriors into twenty bins and normalized the probability density to give a likelihood for each bin, as illustrated in the marginalized timing distributions from Appendix~\ref{ap:posteriors}. We then ran two sets of dynamical fits for each system using either these skewed timing posteriors or a symmetric Gaussian distribution with a width equal to the average of our positive and negative uncertainties.

We fitted dynamical models to the transit timing data using a Differential Evolution Markov Chain Monte Carlo algorithm \citep{t06, n14, jh15, jh16}. We used uniform priors for the orbital period and phase and uniform positive definite priors for the dynamical masses. For each eccentricity vector component, we assumed a Gaussian distribution centered on 0 with a width of 0.1 for the prior. This is wider than the inferred eccentricity distribution among \textit{Kepler}'s multi-planet systems \citep{f14, hl14}, but TTV modeling is subject to an eccentricity-eccentricity degeneracy whereby aligned orbits can have larger eccentricities than allowed by our prior with little effect on the relative eccentricity \citep{jh16}. The results of our dynamical modeling are given in Table \ref{dynamical}. This table includes orbital periods (solved at our chosen epoch of BJD = 2455680), masses, and eccentricity vectors for retrievals with only the \textit{Kepler} data, retrievals including the new WIRC transit time with a Gaussian uncertainty distribution, and retrievals using the skewed WIRC timing posterior. We find that our fits using Gaussian posteriors are generally in good agreement with results from fits utilizing the skewed transit timing posteriors.

\begin{deluxetable*}{rccccc}
\tabletypesize{\scriptsize}
\tablecaption{Results from our dynamical analysis. \label{dynamical}}
\tablehead{\colhead{Planet} & \colhead{Dataset} & \colhead{$P$ [days]} & \colhead{$\Big(\frac{M_\mathrm{p}}{M_\Earth}\Big)\Big(\frac{M_\Sun}{M_\star}\Big)$} & \colhead{$e\cos(\omega)$} & \colhead{$e\sin(\omega)$}}
\startdata
Kepler-29b & Kep LC & $10.33838^{+0.00030}_{-0.00027}$ & \phm{1}4.6$^{+1.4\phm{1}}_{-1.5\phm{1}}$ & -0.060$^{+0.072}_{-0.071}$ & -0.030$^{+0.072}_{-0.072}$ \\
\phm{1} & Kep LC + WIRC (G) & $10.33974^{+0.00014}_{-0.00015}$ &\phm{1}3.7$^{+1.3\phm{1}}_{-1.3\phm{1}}$ & \phm{-}0.013$^{+0.071}_{-0.071}$ & -0.016$^{+0.056}_{-0.063}$ \\
\phm{1}& Kep LC + WIRC (S) & $10.33966 ^{+0.00015}_{-0.00017}$ &\phm{1}3.8$^{+1.1\phm{1}}_{-1.0\phm{1}}$ & \phm{-}0.003$^{+0.068}_{-0.070}$ & -0.088$^{+0.059}_{-0.058}$ \\
\tableline
Kepler-29c & Kep LC & $13.28843^{+0.00048}_{-0.00053}$& \phm{1}4.07$^{+2.87}_{-2.29}$ & \phm{-}0.007$^{+0.063}_{-0.062}$ & -0.022$^{+0.063}_{-0.063}$ \\
\phm{1} & Kep LC + WIRC (G) &$13.28613^{+0.00026}_{-0.00021}$& \phm{1}3.28$^{+1.06}_{-1.08}$ & -0.023$^{+0.061}_{-0.062}$ & -0.022$^{+0.045}_{-0.055}$ \\
\phm{1} & Kep LC + WIRC (S) &$13.28633^{+0.00031}_{-0.00027}$& \phm{1}3.39$^{+0.86}_{-0.84}$ & -0.007$^{+0.059}_{-0.061}$ & -0.085$^{+0.051}_{-0.051}$ \\
\tableline
Kepler-36b & Kep LC & $13.86834^{+0.00050}_{-0.00051}$& \phm{1}3.990$^{+0.093}_{-0.092}$ & \phm{-}0.050$^{+0.023}_{-0.025}$ & -0.026$^{+0.034}_{-0.033}$ \\
\phm{1} & Kep LC + WIRC (G) & $13.86825^{+0.00050}_{-0.00050}$& \phm{1}3.972$^{+0.078}_{-0.074}$ & \phm{-}0.041$^{+0.019}_{-0.020}$ & -0.011$^{+0.018}_{-0.018}$ \\
\phm{1}& Kep LC + WIRC (S) & $13.86821^{+0.00049}_{-0.00049}$& \phm{1}3.964$^{+0.077}_{-0.068}$ & \phm{-}0.037$^{+0.019}_{-0.018}$ & -0.004$^{+0.012}_{-0.015}$ \\
\tableline
Kepler-36c & Kep LC & $16.21867^{+0.00010}_{-0.00010}$& \phm{1}7.456$^{+0.167}_{-0.168}$ & \phm{-}0.053$^{+0.021}_{-0.023}$ & -0.039$^{+0.031}_{-0.031}$\\
\phm{1} & Kep LC + WIRC (G) & $16.21865^{+0.00010}_{-0.00010}$& \phm{1}7.397$^{+0.104}_{-0.107}$ & \phm{-}0.046$^{+0.017}_{-0.018}$ & -0.026$^{+0.017}_{-0.017}$\\
\phm{1} & Kep LC + WIRC (S) & $16.21865^{+0.00010}_{-0.00010}$& \phm{1}7.371$^{+0.092}_{-0.093}$ & \phm{-}0.042$^{+0.017}_{-0.016}$ & -0.019$^{+0.012}_{-0.014}$\\
\tableline
 KOI-1783.01& Kep LC & $134.4622^{+0.0035}_{-0.0038}$& 90.2$^{+30.3}_{-23.2}$ & \phm{-}0.0079$^{+0.0080}_{-0.0050}$ & -0.039$^{+0.012}_{-0.021}$ \\
\phm{1} & Kep LC + WIRC (G) &$134.4628^{+0.0033}_{-0.0035}$& 78.1$^{+15.1}_{-12.9}$ & \phm{-}0.0073$^{+0.0067}_{-0.0046}$ & -0.048$^{+0.014}_{-0.015}$ \\
\phm{1} & Kep LC + WIRC (S) & $134.4629^{+0.0033}_{-0.0036}$& 76.4$^{+11.8}_{-9.6\phm{1}}$ & \phm{-}0.0072$^{+0.0067}_{-0.0045}$ & -0.049$^{+0.014}_{-0.012}$ \\
\tableline
 KOI-1783.02 & Kep LC &  $284.230^{+0.044}_{-0.031}$& 17.1$^{+5.1\phm{1}}_{-4.3\phm{1}}$ & \phm{-}0.018$^{+0.018}_{-0.015}$& -0.011$^{+0.027}_{-0.032}$ \\
\phm{1}  & Kep LC + WIRC (G) & $284.215^{+0.026}_{-0.021}$& 16.2$^{+4.7\phm{1}}_{-3.8\phm{1}}$ & \phm{-}0.017$^{+0.015}_{-0.015}$& -0.020$^{+0.034}_{-0.028}$ \\
 \phm{1} & Kep LC + WIRC (S) & $284.212^{+0.024}_{-0.018}$& 16.1$^{+4.6\phm{1}}_{-3.8\phm{1}}$ & \phm{-}0.017$^{+0.015}_{-0.014}$& -0.020$^{+0.034}_{-0.026}$ \\
 \tableline
 Kepler-177b & Kep LC  & $35.8591^{+0.0019}_{-0.0017}$& \phm{1}5.76$^{+0.84}_{-0.81}$ & -0.026$^{+0.074}_{-0.075}$ & -0.014$^{+0.065}_{-0.068}$  \\
\phm{1} & Kep LC + WIRC (G) & $35.8601^{+0.0015}_{-0.0014}$& \phm{1}5.44$^{+0.78}_{-0.75}$ & \phm{-}0.017$^{+0.052}_{-0.054}$ & -0.001$^{+0.062}_{-0.063}$  \\
 \phm{1} & Kep LC + WIRC (S) & $35.8601^{+0.0013}_{-0.0012}$& \phm{1}5.38$^{+0.78}_{-0.74}$ & \phm{-}0.020$^{+0.047}_{-0.048}$ & \phm{-}0.005$^{+0.061}_{-0.061}$ \\ \tableline
 Kepler-177c & Kep LC & $49.40964^{+0.00097}_{-0.00097}$&  14.6$^{+2.7\phm{1}}_{-2.5\phm{1}}$ & -0.027$^{+0.064}_{-0.065}$ & -0.014$^{+0.056}_{-0.059}$ \\
\phm{1}  & Kep LC + WIRC (G) & $49.40926^{+0.00078}_{-0.00077}$& 13.9$^{+2.7\phm{1}}_{-2.5\phm{1}}$ & \phm{-}0.010$^{+0.045}_{-0.046}$ & -0.003$^{+0.053}_{-0.054}$ \\
 \phm{1} & Kep LC + WIRC (S) &$49.40921^{+0.00072}_{-0.00074}$&  13.5$^{+2.5\phm{1}}_{-2.3\phm{1}}$ & \phm{-}0.013$^{+0.040}_{-0.041}$ & \phm{-}0.003$^{+0.052}_{-0.053}$ \\
 \enddata
 \tablecomments{In the Dataset column, ``Kep LC'' refers to the transit timings from the \textit{Kepler} long-cadence light curves, ``WIRC (G)'' refers to the transit timing from our observations when assumed to have Gaussian uncertainties, and ``WIRC (S)'' refers to the transit timing from our observations taking into account the skewed shape of our timing posteriors. Also, the orbital period $P$ is solved for at our chosen epoch of BJD = 2455680.}
\end{deluxetable*}

\begin{deluxetable}{rcccc}
\tabletypesize{\scriptsize}
\tablecaption{Physical parameters for the planets in this study. \label{densities}}
\tablehead{\colhead{Planet}  & \colhead{$M_\mathrm{p}$ [$M_\Earth$]\tablenotemark{a}} & \colhead{$R_\mathrm{p}$ [$R_\Earth$]\tablenotemark{b}}& \colhead{$\rho_\mathrm{p}$ [g/cm$^3$]} & \colhead{$F_\mathrm{in} [F_\Earth]$\tablenotemark{c}}}
\startdata
Kepler-29b\phn & \phm{1}5.0$^{+1.5\phm{1}}_{-1.3\phm{1}}$ & 2.55$^{+0.12}_{-0.12}$ & 1.65$^{+0.53}_{-0.49}$ & 55.9$^{+6.5}_{-4.8}$ \\
Kepler-29c\tablenotemark{d} & \phm{1}4.5$^{+1.1\phm{1}}_{-1.1\phm{1}}$ &  2.34$^{+0.12}_{-0.11}$ & 1.91$^{+0.57}_{-0.54}$ & 34.4$^{+3.8}_{-3.8}$  \\
Kepler-36b\tablenotemark{d} & \phm{1}3.83$^{+0.11\phm{1}}_{-0.10\phm{1}}$ &  1.498$^{+0.061}_{-0.049}$ & 6.26$^{+0.79}_{-0.64}$ & 247$^{+32}_{-32}$  \\
Kepler-36c\phn  &  \phm{1}7.13$^{+0.18\phm{1}}_{-0.18\phm{1}}$ &  3.679$^{+0.096}_{-0.091}$ & 0.787$^{+0.065}_{-0.062}$ & 191.0$^{+9.7}_{-10.4}$  \\
 KOI-1783.01\phn & 71.0$^{+11.2}_{-9.2\phm{1}}$ &  8.86$^{+0.25}_{-0.24}$ & 0.560$^{+0.101}_{-0.085}$ & \phm{1}5.70$^{+0.27}_{-0.27}$ \\
 KOI-1783.02\tablenotemark{d} & 15.0$^{+4.3\phm{1}}_{-3.6\phm{1}}$ &  5.44$^{+0.52}_{-0.30}$ & 0.51$^{+0.21}_{-0.15}$ & \phm{1}2.49$^{+0.35}_{-0.35}$ \\
 Kepler-177b\tablenotemark{d} & \phm{1}5.84$^{+0.86\phm{1}}_{-0.82\phm{1}}$ &  3.50$^{+0.19}_{-0.15}$ & 0.75$^{+0.16}_{-0.14}$  & 30.4$^{+4.0}_{-4.0}$  \\
 Kepler-177c\phn & 14.7$^{+2.7\phm{1}}_{-2.5\phm{1}}$ &  8.73$^{+0.36}_{-0.34}$ & 0.121$^{+0.027}_{-0.025}$  & 25.4$^{+1.6}_{-1.6}$\\
 \enddata
   \tablenotetext{a}{Calculated from our dynamical masses and the stellar masses of \citet{fp18}.}
   \tablenotetext{b}{Calculated from either our measured $R_\mathrm{p}/R_\star$ or that from \citet{t18} and stellar radii from \citet{fp18}.}
   \tablenotetext{c}{Calculated in the low-eccentricity ($e^2 << 1$) approximation via $F_\mathrm{in}~=~4.62\times10^4F_\Earth\big(\frac{T_\mathrm{eff}}{T_\Sun}\big)^4\big(\frac{a}{R_\star}\big)^{-2}$ \citep{jh16}, with effective temperatures from \citet{f17b} and scaled semi-major axes from our measurements or \citet{t18}.}
  \tablenotetext{d}{Radius ratio and scaled semi-major axis taken from \citet{t18}.}
  
\end{deluxetable}

\section{Results} \label{sec:results}
We determine the significance of each detection in the WIRC data by re-running the joint fit and allowing the WIRC transit depth to vary independent of the \textit{Kepler} transit depth. The confidence is then estimated using the width of the posterior on the WIRC transit depth. We detect transit signals for all four of our targets with $3\sigma$ or greater confidence in the WIRC data alone.

We show various quality statistics for each night of photometry in Table~\ref{photqual} (see Section~\ref{sec:performance}~for additional details). Our results for the photometric fits to each observed planet are given in Table \ref{fitting}, and the resulting orbital periods, masses, and eccentricity vectors are presented in Table \ref{dynamical}. We combine our photometric and dynamical results with previously computed stellar parameters to yield the physical planet parameters we report in Table \ref{densities}. Below we discuss WIRC's overall photometric performance as well as results for each individual system.
\subsection{Instrument Performance \label{sec:performance}}

Our best photometric performance is for Kepler-177c, where we were only $\sim20\%$ above the shot noise. We also investigate how well WIRC mitigates time-correlated noise, which can lead to underestimated uncertainties in reported transit times. We calculate the RMS versus bin size for each observation and show the corresponding plots in the bottom right panels of Figures \ref{kep29fit}--\ref{kep177fit}.
We find that Kepler-29b and KOI-1783.01 appear to have minimal time-correlated noise (see the bottom right panels in Figures \ref{kep29fit} and \ref{koi1783fit}, respectively). Kepler-36c has some time-correlated trends on longer timescales, and for Kepler-177c, quasi-periodic noise is readily visible in both the best-fit residual plot and in the RMS versus bin size plot (see also the bottom right panel in Figures \ref{kep36fit} and \ref{kep177fit}, respectively). We tried adding sinusoids to our fits for these planets, but found that this had a negligible effect on the overall quality of the fits and the resulting transit timing posteriors. 

\begin{figure*}[hbt!]
    \centering
    \includegraphics[width=0.8\textwidth]{{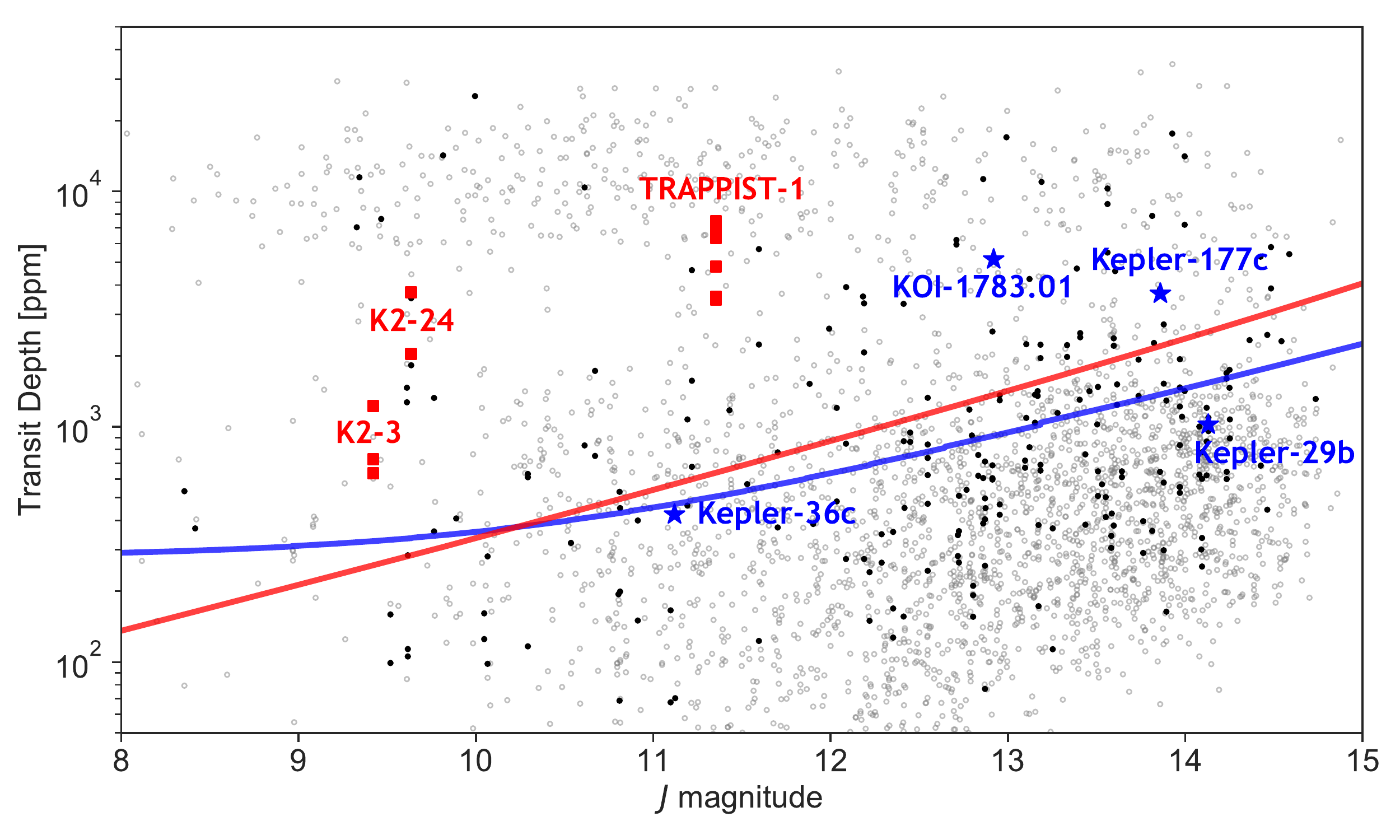}}
    \caption{Transit depth as a function of host star magnitude for non-TTV (grey points) and TTV (black points) systems, taken from the NASA Exoplanet Archive. Also noted are approximate 3$\sigma$ detection thresholds with \textit{Spitzer} (red curve), which is scaled with magnitude from the photometric scatter obtained by \citet{b17} with a slight nonlinear correction at higher magnitudes fit to the brown dwarf survey results of \citet{m15b}, and the 3$\sigma$ detection threshold with WIRC assuming the optimal co-addition strategy (blue curve). The systems investigated in this work are marked with labeled blue stars, while a few sample TTV systems investigated by \textit{Spitzer} (K2-3, K2-24, TRAPPIST-1) are given marked with labeled red squares \citep{b16, d18, p18}. The WIRC detection threshold levels off for brighter stars due to decreasing observing efficiency, and the slight discontinuities in the curve are artifacts of discrete changes in the number of co-additions.}
    \label{betterthanspitzer}
\end{figure*}

To derive a representative noise statistic for WIRC, we first calculated the scatter in 10 minute bins for each of our observations. These statistics were then scaled to the equivalent values for observations of a 14th magnitude star. In some of our earliest observations we used a sub-optimal co-addition strategy, resulting in relatively inefficient observations (for Kepler-36c, this increased the noise by 31.1\% relative to a more optimal strategy).  We therefore applied an additional correction factor to to rescale the noise for these inefficient observations to the expected value for better-optimized observations. Averaging these corrected noise statistics together, we find that WIRC can deliver 1613~ppm photometry per 10 minute bin on a \textit{J} = 14 magnitude star. If we assume that we are able to collect two hours of data in transit and two hours out of transit, this equates to a precision of 659~ppm on the transit depth measurement for planets around a \textit{J} = 14 magnitude star. To highlight the range of parameter space that this precision opens up, we plot transit depths for all confirmed transiting exoplanets against host star $J$ magnitude in Figure~\ref{betterthanspitzer} along with the $3\sigma$ detection thresholds of WIRC and \textit{Spitzer}. While \textit{Spitzer} performs better for brighter stars, WIRC begins to out-perform \textit{Spitzer} for stars fainter than $\sim10$ magnitude, doing a factor of 1.6 better at \textit{J} = 14. In practice, the achieved photometric precision will also depend on factors such as atmospheric background, amount of baseline obtained, diurnal constraints, and the number of available comparison stars of comparable magnitude, but the first-order considerations in Figure~\ref{betterthanspitzer} suggest that ground-based, diffuser-assisted infrared photometry can indeed outperform some current space-based facilities for typical \textit{Kepler} transiting planet systems.

\subsection{Kepler-29}
Kepler-29b is sub-Neptune near the 5:4 and 9:7 mean-motion resonances with the sub-Neptune Kepler-29c. Both low-density planets were originally confirmed by \citet{f12b} using TTVs; subsequent dynamical analyses have shown that the pair may actually be in the second-order 9:7 resonance \citep{m17b}, but the TTV curve is likely also affected by proximity to the first-order 5:4 resonance \citep{jh16}. We detect a transit of Kepler-29b at $3.5\sigma$ confidence in the WIRC data. The final detrended \textit{Kepler} and WIRC light curves, models, residuals, and RMS binning plots for Kepler-29b are shown in Figure~\ref{kep29fit} and the corresponding posterior probability distributions are shown in Figure~\ref{kep29corner}. Although the transit shape is poorly constrained by the WIRC data alone, both ingress and egress are visible by eye in the WIRC light curve and the relative timing of these two events provides a solid estimate of the transit time when we constrain the transit shape using the \textit{Kepler} photometry. We find that the resulting posterior distribution for our new WIRC transit time is fairly asymmetric, with the final timing offset determined to $-14^{+17}_{-3}$ min.

Our new observation was obtained in an epoch where the Kepler-only dynamical fits yield substantially divergent transit times, and as a result our new transit time provides an improved constraint on the planet masses and eccentricities as shown in Figure~\ref{kep29data}. We find that the dynamical mass estimate for Kepler-29c has improved by almost a factor of three in our updated fits. Our new results favor dynamical masses on the low side of (but not incompatible with) the mass distributions inferred by \citet{jh16} for Kepler-29b and c. 

Despite these decreased masses our updated densities for these planets (1.7$\pm0.5$ and 1.9$\pm0.5$ g/cm$^3$, respectively) are larger than the densities reported by \citet{jh16}. This is because we utilize updated stellar parameters of $M = 0.761^{+0.024}_{-0.028}~M_\Sun$ and $R = 0.732^{+0.033}_{-0.031}~R_\Sun$ from \citet{fp18}, which are smaller than the values of $M = 0.979\pm0.052~M_\Sun$ and $R = 0.932\pm0.060~M_\Sun$ adopted by \citet{jh16}. For a fixed planet-star radius ratio, a smaller stellar radius implies a correspondingly smaller planet radius. Similarly, a smaller stellar mass implies a larger planet mass for the same best-fit dynamical mass ratio. Both changes therefore act to increase the measured planetary density. Even with these increased density estimates, it is likely that both of these planets have retained a modest hydrogen-rich atmosphere (see \S\ref{sec:smalldiscussion}).
The masses and radii of both planets also remain quite similar, in good agreement with the ``peas in a pod'' trend wherein multi-planet \textit{Kepler} systems tend to host planets that are similar in both size and bulk density \citep{m17, w18b}. 
\subsection{Kepler-36}
 The Kepler-36 system includes two planets with strikingly dissimilar densities: Kepler-36b is a rocky super-Earth close to 7:6 mean-motion resonance with the low-density sub-Neptune Kepler-36c \citep{c12}. The latter planet was included in our sample, and we detect it with a significance of 5.3$\sigma$. We present the final light curves and associated statistics for our new transit observation of Kepler-36c in Figure~\ref{kep36fit}, and plot the corresponding posteriors in Figure~\ref{kep36corner}. The posterior distribution on the WIRC transit time is again fairly asymmetric, with the offset constrained to -18$^{+12}_{-5}$ minutes. We obtain masses and densities for both planets consistent with previous investigations \citep[though on the low side for Kepler-36b;][]{c12, hl17}. In Figure~\ref{kep36data}, we provide updated dynamical masses, eccentricity vectors, and transit timing for this system. Future constraints from \textit{TESS} should allow for improved mass estimates in this system, especially for Kepler-36c \citep{g19}.
 
 The RMS scatter achieved for this measurement was $2\times$ the photon noise limit (see bottom right panel of Figure~\ref{kep36fit}), which is higher than any of the other observations presented in this work. This is due in part to scintillation noise  \citep{s17}, as Kepler-36 was our brightest target and we used correspondingly short integration times. For this star, the scintillation noise at an airmass of 1.5 is $\sim650$ ppm, which is comparable to the shot noise. Our use of short integration times also limited our observing efficiency, resulting in higher photometric scatter than might otherwise have been expected for this relatively bright star. Both problems could be mitigated by increasing the number of co-adds, resulting in a longer effective integration time and higher overall observing efficiency.

\subsection{KOI-1783}
As we will discuss in \S\ref{sec:confirmation}, there is already compelling evidence in the literature establishing the planetary nature of this system, which contains two long-period (134 and 284 days, respectively) gas giant planet candidates located near a 2:1 period commensurability. We present the final light curves and associated statistics for our new transit observation of KOI-1783.01 in Figure~\ref{koi1783fit}, and plot the corresponding posteriors in Figure~\ref{koi1783corner}. This planet is detected with a significance of $5.9\sigma$ in the WIRC data, and we achieve a timing precision of about 10 minutes. These results are in good agreement with a model of the KOI-1783 system that assumes the source of TTVs to be near-resonant planet-planet perturbations.
In Figure~\ref{koi1783data}, we present updated constraints on dynamical masses, eccentricities, and transit timing for KOI-1783. Our new transit observation reduces the uncertainty on the dynamical mass of KOI-1783.01 by approximately a factor of two. When combined with the stellar parameters from \citet{fp18}, these new constraints provide the most detailed picture of this system to date. We find that KOI-1783.01 is slightly smaller than Saturn, with $R_\mathrm{p} = 8.9^{+0.3}_{-0.2}R_\Earth$ and $M_\mathrm{p} = 71^{+11}_{-9}M_\Earth$. This corresponds to a density of $\rho = 0.56^{+0.10}_{-0.09}$~g/cm$^3$, consistent with the presence of a substantial gaseous envelope; we discuss the corresponding implications for this planet's bulk composition in more detail in \S\ref{sec:bulkmetallicity}. 
KOI 1783.02 has a mass of $M_\mathrm{p} = 15^{+4}_{-4} M_\Earth$, a radius of $R_\mathrm{p} = 5.4^{+0.5}_{-0.3}R_\Earth$, and a density of $\rho = 0.5^{+0.2}_{-0.2}$g/cm$^3$, again indicative of a substantial gaseous envelope. Both planets appear to have low orbital eccentricities ($e\lesssim0.05$), in agreement with the overall \textit{Kepler} TTV sample \citep{f14, hl14, x16}. Additionally, we note that the uncertainty on $e\cos(\omega)$ for KOI-1783.01 is an order of magnitude lower than for the other planets in this study, corresponding to a $\pm1\sigma$ uncertainty of approximately 13 hours in the secondary eclipse phase. Although this is quite good for a planet on a 134 day orbit, the star's faintness and the planet's low equilibrium temperature make this a challenging target for secondary eclipse observations.
\subsection{Kepler-177}
The Kepler-177 system contains a low-density sub-Neptune (Kepler-177b) and a very-low-density sub-Neptune (Kepler-177c) located near the 4:3 mean motion resonance.  This system was initially confirmed via TTVs by \citet{x14} and subsequently re-analyzed by \citet{jh16} and \citet{hl17}. Our final light curves and associated statistics for Kepler-177c are given in Figure~\ref{kep177fit}, and the posteriors are given in Figure~\ref{kep177corner}. We detect the transit at $5.5\sigma$ significance and measure the corresponding transit time with a $1\sigma$ uncertainty of approximately 10 minutes. Although our new dynamical fits for this system result in modestly lower mass uncertainties, our transit observation was taken close to one TTV super-period away from the \textit{Kepler} data, where diverging solutions re-converge and thus our new observations provided limited leverage to constrain these dynamical models. If the \textit{TESS} mission is extended it should provide additional transit observations that would further reduce the mass uncertainties in this system \citep{g19}, but our observations demonstrate that this system is also accessible to ground-based follow-up at a more favorable epoch.
\section{Discussion \label{sec:discussion}}
\subsection{Confirmation of the KOI-1783 System \label{sec:confirmation}}
\begin{figure*}[ht!]
    \centering
    \includegraphics[width=\textwidth]{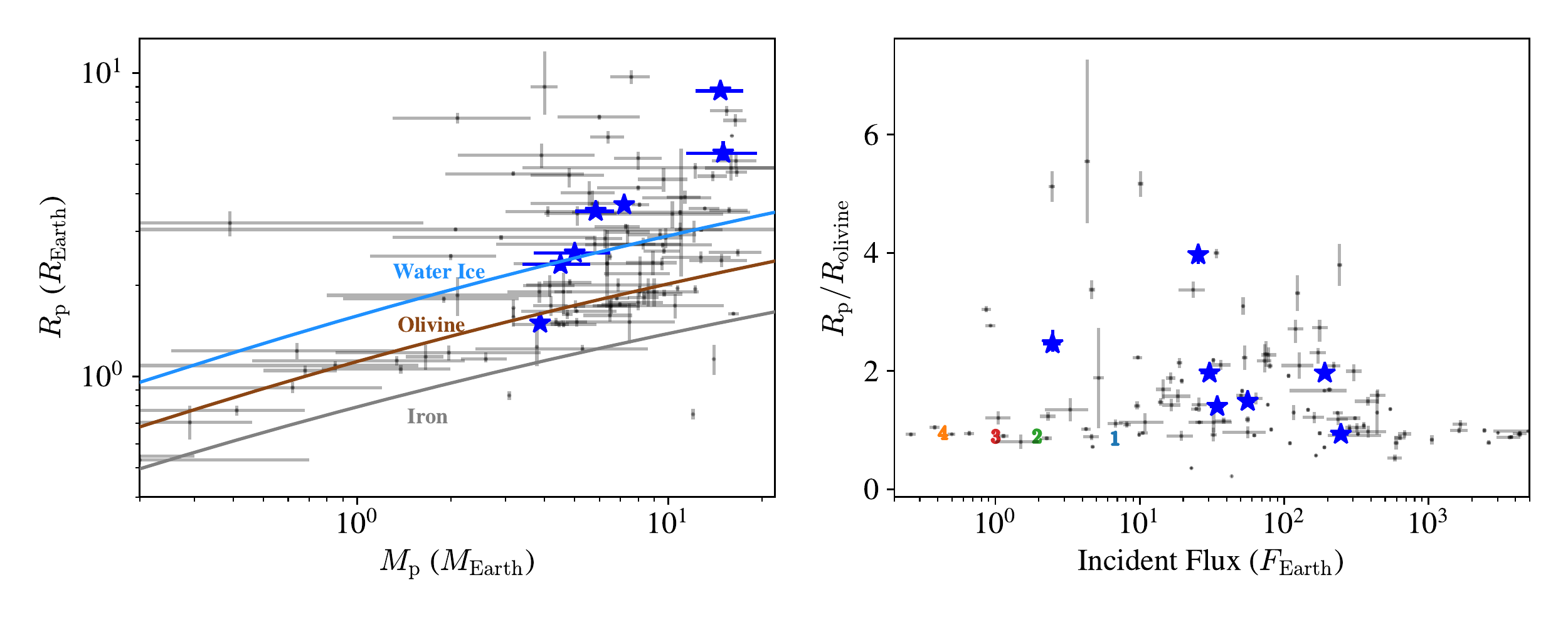}
    \caption{(Left) Masses and radii of the sub-Neptune planets studied in this work (blue stars) compared to all $M < 20M_\Earth$ planets from the NASA Exoplanet Archive (gray points). The blue, brown, and grey curves show the mass-radius relation for planets made of pure water ice, olivine, and iron \citep{f07}. (Right) Planetary radius relative to that of a pure-rock planet of the same mass is plotted as a function of incident flux for our systems (blue stars) and  all $M < 17M_\Earth$ planets on the NASA Exoplanet Archive (gray points). Also noted are the Solar System planets with the colored numbers (Mercury is 1, Venus is 2, Earth is 3, and Mars is 4).}
    \label{smallplanets}
\end{figure*}
As the only unverified planet candidate in our sample, KOI 1783.01 represents a special case for this program. A transiting planet candidate around KOI-1783 (KIC 10005758) was first reported by \citet{b13}, and a second candidate in the system was identified by the Planet Hunters citizen science collaboration \citep{l13}. While the \textit{a priori} probability of both transit signals being false positives is quite low \citep{l11, l12, l13, l14a, r14a}, a few characteristics of this system precluded a quick confirmation. First, the transit signals for both candidates are near-grazing (the grazing parameter $X~=~b~+~R_\mathrm{p}/R_\star$ is 0.9949$^{+0.0032}_{-0.0027}$ for KOI-1783.01 from our posteriors, and 0.932$^{+0.065}_{-0.015}$ for KOI-1783.02 from the \citet{t18} catalog), with ``V''-shaped morphologies that \citet{b13} noted as being potentially diagnostic of an eclipsing binary. Additionally, the \textit{Kepler} Data Validation reports show a fairly large offset ($\sim0\arcsec.25$) of the stellar centroid during the transit relative to the KIC position, which is also typical of stellar blends. 

The two transit candidates in this system have a period ratio of 2.11, near the 2:1 commensurability. Such an architecture can generate detectable TTVs, which previous studies have used to confirm the planetary nature of transit candidates \citep{s13, n13}. Early analyses of the transit times of KOI-1783.01 \citep{f12a, m13} noted the potential presence of TTVs, but concluded that the significance of the deviation from a linear ephemeris was too low to be conclusive. As \textit{Kepler} continued to observe this target, evidence for TTVs of both planet candidates in this system grew stronger \citep{r14a, h16}. An independent analysis of this system by the Hunt for Exomoons with Kepler Project found evidence for dynamical interactions \citep{k15a}, selecting a TTV model over a linear ephemeris model by 17.2$\sigma$ for KOI-1783.02. The spectral TTV analysis of \citet{o18} also found evidence of dynamical interactions, yielding $\Delta\chi^2$ values for the TTV signals over a linear model of 49 and 264 for KOI-1783.01 and .02, respectively (the authors note that $\Delta\chi^2 \gtrsim 20$ is a reliable detection threshold). 

For non-dynamically interacting systems, it is common to use statistical arguments to establish that the planetary hypothesis is the most likely explanation for a given transit signal using codes such as the publicly-available false-positive probability (FPP) calculator \texttt{vespa} \citep{m12, m15a}. The \texttt{vespa} package has been used to statistically validate more than a thousand exoplanet candidates from \textit{Kepler} and \textit{K2} thus far \citep{c16, m16, l18a, l18b, m18}, although refutation of some previously validated planets suggests that caution is necessary when validating with limited follow-up data \citep{s16b, c17a, s17b}. \citet{m16} obtained FPPs for all KOIs, including KOI-1783.01 (FPP = $0.680 \pm 0.014$) and KOI-1783.02 (FPP = $0.200 \pm 0.012$). However, TTVs were not considered in the construction of the light curves for these planets, which can inflate the FPP by making the transits look more ``V''-shaped. Additionally, \citet{m16} found four confirmed planets with anomalously high FPPs: three exhibited TTVs, and the other had grazing transits. Our analysis suggests that KOI-1783 system is a near-grazing TTV system, making it very likely to have an overestimated FPP.

In a six-year campaign, \citet{s16b} performed RV observations of a sample of 125 KOI stars, including KOI-1783. They observed KOI-1783 two times with SOPHIE and detected no RV variation. Additionally, they establish 99\% upper limits on the RV semi-amplitude ($K < 81.3$ m/s) and corresponding mass ($M < 2.83\ M_\mathrm{J}$). While these upper limits were derived by fitting a circular orbit with no TTVs, the lack of detected RV variations rule out the eclipsing binary false positive mode to very high confidence. 

In addition to high-resolution spectroscopic follow-up, three ground-based adaptive optics (AO) follow-up observations of KOI-1783 have been performed to date, as listed by \citet{f17a} and the Exoplanet Follow-up Observing Program. The Robo-AO team observed this star in their LP600 filter with the Palomar 60" telescope, achieving a contrast of $\Delta M = 4.00$ mag at $0\arcsec.30$ \citep{l14b}. Additionally, \citet{w15} observed KOI-1783 in $K_s$ band with PHARO on the Hale 200" telescope at Palomar Observatory, achieving a contrast of $\Delta M = 4.33$ mag at $0\arcsec.50$. More stringent contrast constraints of $\Delta M = 7.96$ mag at $0\arcsec.50$ were obtained with NIRC2 on the Keck II Telescope using the Br$\gamma$ filter \citep{f17a}. These observations demonstrate that there are no nearby stars that might explain the $0\arcsec.25$ offset noted in the Data Validation Report.

Published RV data rule out the existence of an eclipsing binary, and AO imaging data rule out the existence of companions. Combined with the aforementioned multiple independent analyses all supporting dynamical interactions between the bodies in the system, these follow-up constraints lead us to conclude that the two transit candidates in the KOI-1783 system should be confirmed as bona fide planets.
\subsection{Population-Level Trends}
\subsubsection{TTVs Probe Warm Sub-Neptune-Sized Planets 
\label{sec:smalldiscussion}}
There are currently very few sub-Neptune-sized transiting planets with well-measured masses at large orbital distances ($P > 100$ days); these systems are quite rare to begin with, and most are too small and faint to be amenable to RV follow-up \citep{jh19}. TTV studies that probe this regime are thus quite valuable, as planets that receive low incident fluxes are much more likely to retain their primordial atmospheres than their more highly-irradiated counterparts \citep[e.g.][]{ow13, m16b}. Even if mass loss is common for these longer-period planets, the mechanism by which it occurs may be quite different. For highly irradiated exoplanets, atmospheric mass loss is primarily driven by thermal escape processes as the intense XUV flux heats the upper atmospheres \citep[e.g.][]{o19}. However for planets on more distant orbits, non-thermal processes are competitive with or dominant over photoevaporative escape; this is, for instance, the present case for terrestrial planets like Mars \citep{t13, t15}. Density constraints for this population of long-period extrasolar planets at low ($\lesssim100F_\Earth$) incident fluxes are therefore critical for building a holistic understanding of atmospheric mass loss in the regime relevant for potentially habitable terrestrial planets.  

In Figure~\ref{smallplanets}, we plot the masses and radii of our sub-Neptune-sized sample ($M < 17M_\Earth$) along with those from the NASA Exoplanet Archive and compare their radii to their incident fluxes. Other than the rocky super-Earth Kepler-36b \citep{c12}, all of the planets in our sample are more inflated than they would be if they were purely composed of silicate rock \citep{f07}, implying that they possess at least modest volatile-rich envelopes. Even after allowing for water-rich compositions, our bulk density estimates for the planets in Table~\ref{densities} are still too low, and likely require a modest hydrogen-rich atmosphere. For Kepler-29b, Kepler-29c, Kepler-36c, and Kepler-177b, the grids of \citet{lf14} suggest hydrogen-helium envelope fractions of 2-5\% in mass. For the more massive sub-Neptunes KOI-1783.02 and Kepler-177c, these grids suggest hydrogen-helium envelope fractions greater than 10\% in mass. In the following section, we explore the bulk composition of KOI-1783.01, KOI-1783.02, and Kepler-177c in more detail.

\subsubsection{Bulk Metallicities of the Giant Planets KOI-1783.01, KOI-1783.02, and Kepler-177c}
\label{sec:bulkmetallicity}
TTVs can also deliver masses and radii for giant planets in the low-insolation regime. This is crucial for estimates of bulk metallicity, as gas giants hotter than approximately 1000 K appear to have inflated radii that are inconsistent with predictions from standard interior models  \citep[e.g.,][]{l11b, t16a, t18b}. Relatively cool, dynamically interacting planets such as KOI-1783.01 are not expected to be affected by this inflation mechanism and are therefore ideal candidates for these studies. 

\begin{figure}[htb!]
    \centering
    \includegraphics[width=0.47\textwidth]{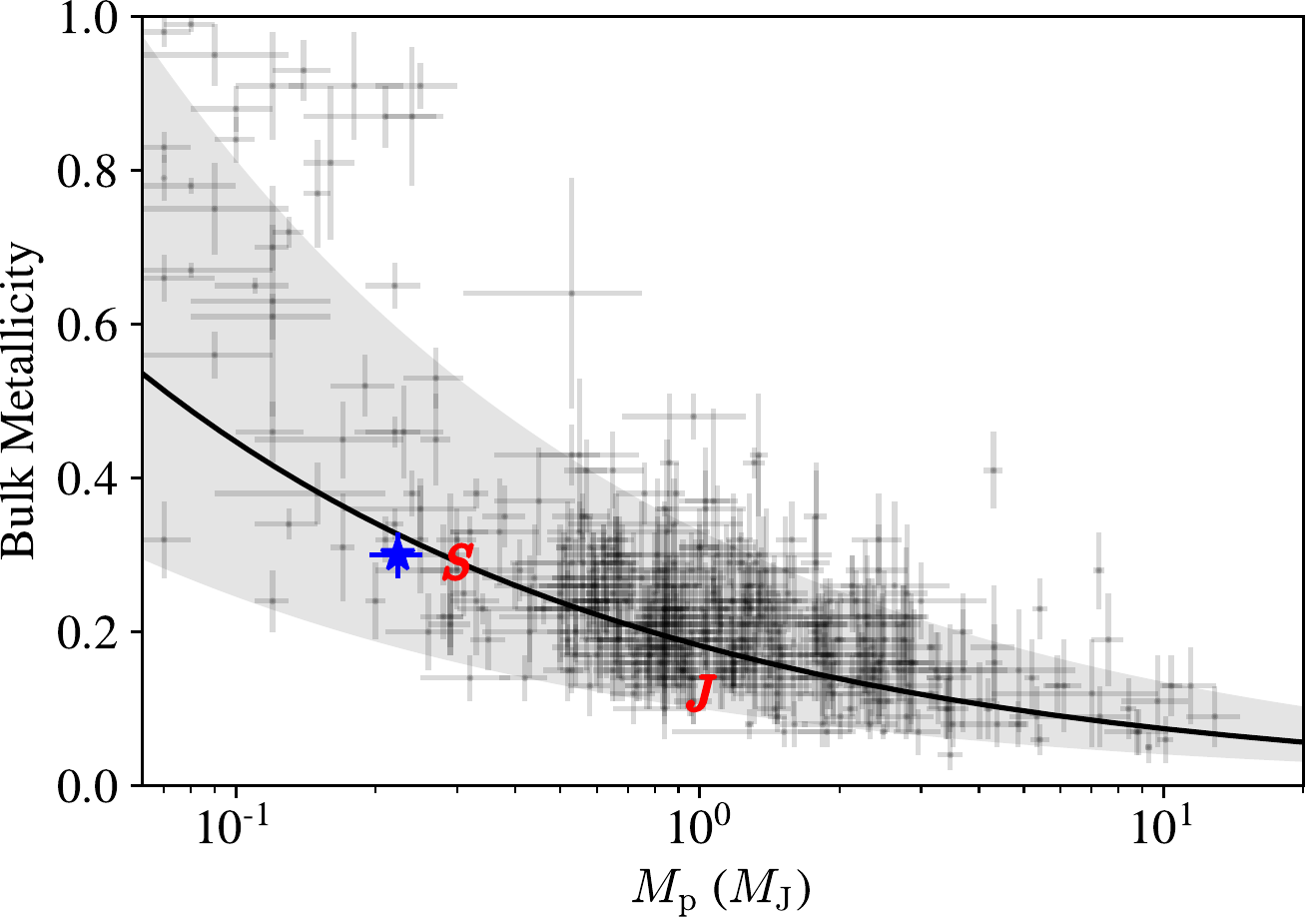}
    \caption{Bulk metallicity of KOI-1783.01 (blue star) compared to the metallicities of the \citet{tf19} sample (grey points). The best-fit mass-metallicity relation obtained by \citet{t16a} is shown in black, with $\pm1\sigma$ uncertainties denoted by the grey shaded region. The red ``J'' and ``S'' correspond to Jupiter and Saturn.}
    \label{bigplanets}
\end{figure}

We measure the mass of the gas giant KOI-1783.01 to $\sim15\%$ precision and its radius to $\sim3\%$, as this star has relatively accurate stellar parameters from \citet{fp18}. When combined with our incident flux constraints and stellar age estimates from \citet{fp18}, these parameters yield a bulk metallicity of $Z_\mathrm{p} = 0.30\pm0.03$ for KOI-1783.01 using the statistical model of \citet{tf19}. Using the stellar metallicity from Table \ref{stellar} and the $Z_\mathrm{star} = 0.014 \times 10^{\mathrm{[Fe/H]}}$ prescription from \citet{t16a}, this corresponds to $Z_\mathrm{p}/Z_\mathrm{star} = 16.6^{+2.4}_{-2.2} $. We note that when masses and radii are constrained to this level of precision we should also consider the additional uncertainties introduced by the choice of models, which are not accounted for in these error bars \citep{t16a,tf19}.  This bulk metallicity value is nevertheless in excellent agreement with the mass-metallicity relation previously inferred for gas giant planets at higher incident fluxes \citep{t16a, tf19}, as shown in Figure~\ref{bigplanets}.

This bulk metallicity also yields an upper limit on the atmospheric metallicity, as the metallicity observable in a planetary atmosphere will always be less than the total metal content of the planet \citep{tf19}. For KOI-1783.01, this (95th percentile) upper limit is $Z_\mathrm{atm} \leq 79\times$ solar, where ``solar'' refers to the \citet{a09} photospheric metal fraction of $1.04\times10^{-3}$. This calculation assumes an average mean molecular mass of 18 (that of water) for this heavy element component; if this is not the case, then the true upper limit on the atmospheric metallicity should be scaled by $18/\mu_Z$ \citep{tf19}.

We calculate comparable bulk composition estimates for the two sub-Neptunes in our sample, KOI-1783.02 and Kepler-177c. In this mass regime, differences in equation of state between rock and water ice become important, adding another degree of freedom to the calculation. We construct models composed of a rock layer, a water layer, and low-density H/He layer enriched to Neptune's metallicity (90$\times$ solar) by borrowing water from the water layer. We do not include mass loss in our simulation, and we assume negligible amounts of iron in the calculation. We use constraints on the mass, radius, host star age, and incident flux to retrieve the composition, including the relative amounts of rock, water, and H/He. Although we are not able to place strong constraints on the relative amounts of rock versus water as the radius is still fairly insensitive to the core composition details \citep{lf14, p17b}, we are able to place a strong constraint on the total bulk metallicity $Z_\mathrm{p}$ and the corresponding the H/He fraction $f_\mathrm{H/He} = 1 - Z_\mathrm{p}$.

As hinted at by their low bulk densities, these two planets have large H/He mass fractions: $f_\mathrm{H/He} = 0.31 \pm0.08$ for KOI-1783.02 and $f_\mathrm{H/He} = 0.74 \pm0.04$ for Kepler-177c. The value for Kepler-177c is somewhat problematic from a planet formation perspective, as it implies a maximum core mass of just 4 $M_\Earth$. Depending on the planet's formation location, it may be difficult to explain how such a small core could have accreted such a massive gas envelope. One explanation is that the core formed outside 1 au and experienced relatively dust-free accretion, as is typically invoked for super-puffs \citep{lc16}. We note, however, that super-puffs are a few times less massive than Kepler-177c despite having similar inferred core masses, implying that the gas-to-core mass ratio of Kepler-177c exceeds that of a typical super-puff. Although it is possible that our estimate of this maximum core mass might have been biased by assumptions made in our models, accounting for atmospheric mass loss would have preferentially removed hydrogen and helium, and including iron in the model would have increased the $f_\mathrm{H/He}$. We conclude that these assumptions are unlikely to explain the large inferred H/He mass fraction for this planet. The MIST isochrone-derived age estimate for this planet from \citet{fp18} appears to be quite secure, with $\log(\mathrm{age}) = 10.07\pm0.04$, so it is unlikely that this planet's radius is inflated by residual heat from formation. 

Can Kepler-177c be inflated by internal heating mechanisms such as Ohmic dissipation \citep{pu17} or obliquity tides \citep{mill19}? Its large total mass and low insolation makes this scenario unlikely. We assess the scenario of Kepler-177c having a core mass of 14.5$M_\oplus$ and an envelope mass of 0.2$M_\oplus$ (envelope mass fraction of 1\%). Its estimated equilibrium temperature is $\sim$800K, too low for Ohmic dissipation to puff up Kepler-177c to $\gtrsim$8$R_\oplus$ \citep[see Figures 8 and 9 of ][]{pu17}. Next, we assess heating by obliquity tides. Even if we assume maximal obliquity, the expected thickness of the envelope is $\sim$0.48$R_\oplus$ \citep[see equation 13 of][]{mill19}. If the composition of Kepler-177c core is similar to that of Earth, we expect its core size to be $\sim$1.95$R_\oplus$ (assuming $R \propto M^{1/4}$), so that the expected total radii of the planet is only $\sim$2.43$R_\oplus$, far too small to explain the measured 8.73$R_\oplus$. Even at gas-to-core mass ratio of 10\%, the expected total radii is just 3.74$R_\oplus$.

\subsection{A Possible Formation Scenario for Kepler-177}
\begin{figure}[htb!]
    \centering
    \includegraphics[width=0.47\textwidth]{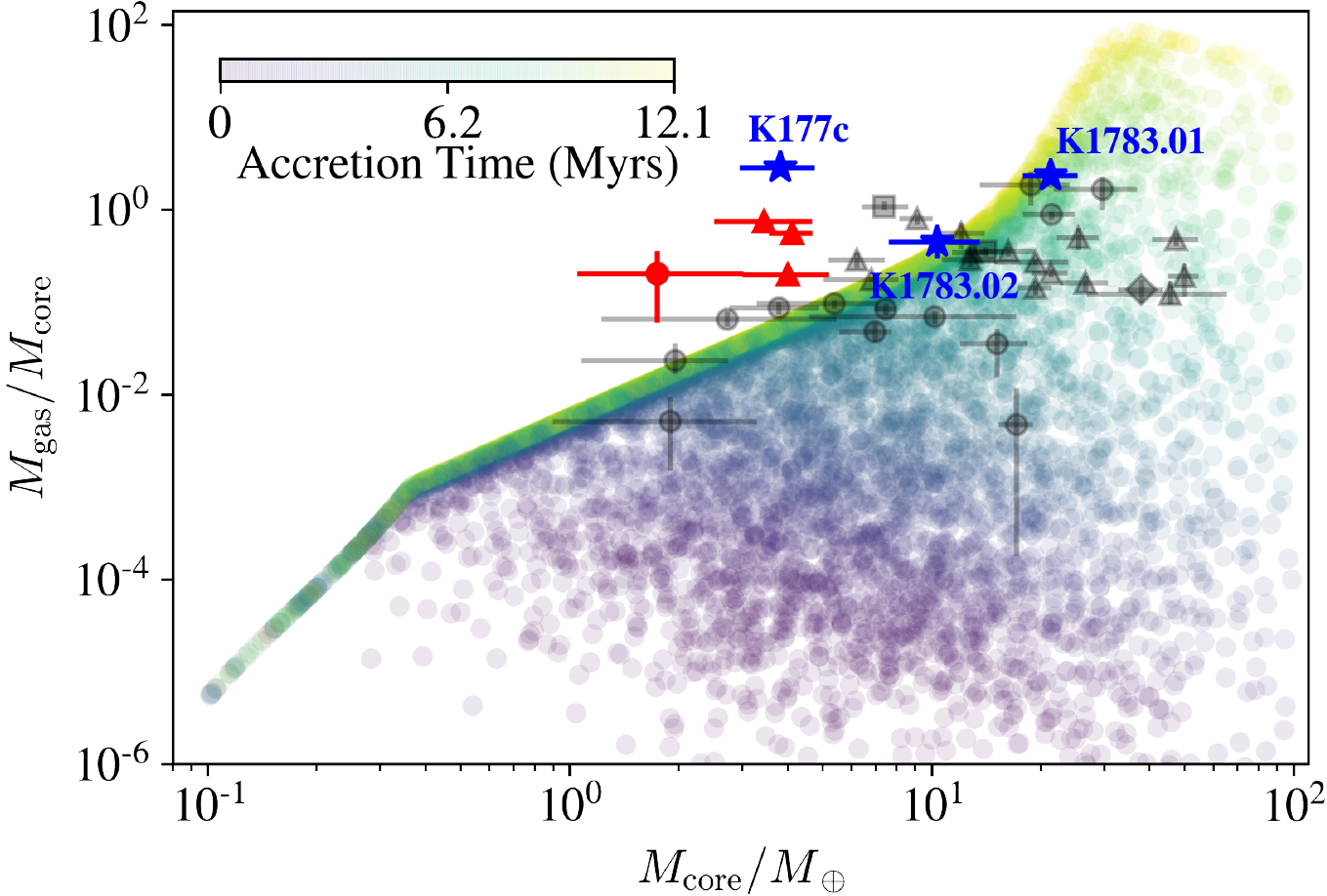}
    \caption{The \citet{l19} gas-to-core mass ratio (GCR) plot as a function of core mass $M_\mathrm{core}$ and accretion time (color-coded) for their best-fit model ensemble of core masses (log-normal with $\mu = 4.3M_\Earth$ and $\sigma = 1.3$). Overplotted on this theoretically-derived distribution are observational GCR constraints on real planets, denoted by gray circles \citep{lf14}, gray triangles \citep{p17b}, gray diamonds \citep{d18b}, gray squares \citep{p18}, and blue stars (this work). Previously identified super-puffs (Kepler-51b, Kepler-223e, Kepler-87c, and Kepler-79d) are marked in red. Note that Kepler-177c has a larger GCR than these super-puffs despite having a similar core mass.}
    \label{eveplot}
\end{figure}

We conclude that Kepler-177c rightfully belongs in the small sample of $\sim15M_\Earth$ planets with extremely low bulk densities (and thus extremely large envelope fractions). This sample also includes Kepler-18d \citep{c11, p17b} and K2-24c \citep{p18}. \citet{p18} suggest a formation scenario for the latter planet wherein the disk dissipates just as the planet begins to enter runaway accretion. \citet{l19} show that the sub-Saturn population can indeed be explained by the timing of disk dispersal, but they note as a prerequisite that their cores must be massive enough to trigger runaway accretion during the disk lifetime, $\gtrsim 10M_\Earth$. For cores less massive than this, the maximum gas-to-core mass ratio (GCR) is set by the amount of gas that can be accreted by cooling. In Figure~\ref{eveplot}, we reproduce the \citet{l19} GCR plot as a function of core mass and accretion time, which highlights the different regimes dictating the maximum envelope fraction for a given core mass. While KOI-1783.01 and KOI-1783.02 can largely be explained within the framework of disk dispersal timing relative to the onset of runaway accretion, Kepler-177c cannot, nor can K2-24c or Kepler-18d. These low-density $15M_\Earth$ planets are outliers, lying above their theoretical maximum GCRs, as are the super-puffs Kepler-51b \citep{m14}, Kepler-223e \citep{m16c}, Kepler-87c \citep{o14}, and Kepler-79d \citep{jh14}.

As a result, \citet{l19} suggests that these more massive low-density planets may share a formation pathway with the less-massive super-puffs. Super-puffs likely accreted their envelopes farther from their star and then migrated inwards \citep{ih12, l14, g16, lc16, s18b}, and additionally should have experienced ``dust-free'' accretion, meaning that dust did not contribute much to the overall opacity due to e.g. grain growth or sedimentation \citep{lc15, lc16}. To test the feasibilty of this hypothesis, we can estimate the amount of time that Kepler-177c must have spent undergoing dust-free accretion and compare to typical disk lifetimes. If this timescale is longer than the typical disk dispersal timescale, then a mechanism other than dust-free accretion is necessary; if it is comparable or shorter, then dust-free accretion may be feasible. For Kepler-177c ($M_\mathrm{core} \approx 3.8M_\Earth, \mathrm{GCR} \approx 2.8$), we can approximate the dust-free accretion time necessary to achieve the observed GCR beyond 1 au in a gas-rich disk using the analytic scaling relation of \citet[][see their Equation 24]{lc15}:

\begin{equation}
    t \sim \mathrm{1\ kyr} \Bigg[\Big(\frac{\mathrm{GCR}}{0.1}\Big)\Big(\frac{5M_\Earth}{M_\mathrm{core}}\Big)\Bigg]^{2.5} \approx 8.2\ \mathrm{Myr},
    \label{eq:tacc_k177c}
\end{equation}
where for simplicity we have assumed their nominal values for the $f$ factor, the nebular gas metallicity $Z$, the adiabatic gradient $\nabla_\mathrm{ad}$, and the temperature and mean molecular weight at the radiative-convective boundary $T_\mathrm{rcb}=200$ K and $\mu_\mathrm{rcb}$. The outer layers of dust-free envelopes are largely isothermal so the adopted temperature corresponds to the nebular temperature at the formation location. The estimated accretion timescale required to build Kepler-177c is comparable to typical disk lifetimes \citep[$\sim$ 5 Myr; see, e.g.][and references therein]{a14}. We note that Equation \ref{eq:tacc_k177c} is derived assuming the self-gravity of the envelope is negligible compared to the gravity of the core. The rate of accretion starts to accelerate once GCR $\gtrsim$ 0.5, so a more careful calculation would provide an even shorter timescale. We suggest that 15$M_\Earth$ planets with large GCRs may indeed share a dust-free accretion history with their lower-mass super-puff counterparts. As such, detailed characterization of Neptune-mass planets with low ($\rho \lesssim 0.3$ g/cm$^3$) bulk densities may provide invaluable insights into super-puff formation processes.

\section{Conclusions and Future Prospects} \label{sec:conc}
We presented infrared photometry for four dynamically interacting \textit{Kepler} systems. With precise telescope guiding and the use of an engineered diffuser, we achieved a precision with WIRC that is comparable to or better than \textit{Spitzer} for stars fainter than $J = 9.5$. Most of the planets we observed have host stars that are too faint for standard Doppler-based follow-up, but their masses can be measured to a high relative precision by fitting their transit timing variations. Our new transit measurements demonstrate that a single, well-timed follow-up observation taken years after the \textit{Kepler} mission's conclusion can improve mass estimates by almost a factor of three. Perhaps unsurprisingly, we found that observing in epochs of maximally divergent transit times for differing dynamical solutions yields the largest improvements in mass estimates. The potential information gain is also larger for long-period systems with relatively few transits observed during the original \textit{Kepler} mission. The systems we have studied highlight the diverse range of science cases made possible by diffuser-assisted photometry, including the confirmation of long-period planet candidates in TTV systems as well as bulk composition studies for relatively cool planets ranging in size from sub-Neptunes to gas giants.

WIRC's demonstrated infrared photometric precision opens up multiple new opportunities for ground-based studies of transiting planets and brown dwarfs. For dynamically interacting systems bright enough for RV observations, diffuser-assisted transit observations can provide an extended TTV baseline for joint RV-TTV modeling.  These kinds of studies can constrain the structures of planetary systems without reliance on stellar models \citep{a15, a16, af17, w17, a18b, p18}. For highly irradiated gas giant planets, WIRC can be used to complement existing space-based emission and transmission spectroscopy from \emph{Spitzer} and the \emph{Hubble Space Telescope} by observing photometric transits and secondary eclipses at wavelengths that are inaccessible to these telescopes.  This extended wavelength coverage is important for reducing degeneracies in atmospheric retrievals \citep[e.g.][]{b12, l12b, l13b, l14c}. WIRC can also measure low-amplitude rotational variability in brown dwarfs at infrared wavelengths. Current ground-based infrared measurements can constrain variability at the $\sim 0.7\%$ level \citep{w14, r14b} in these objects; for the brighter ($J$ = 14-15) variable brown dwarfs, WIRC will be able to push these limiting amplitudes below $0.1\%$. We are only beginning to explore the parameter space made available by diffuser-assisted photometry, but the prospects for new ground-based studies of brown dwarfs and transiting planets are promising.

\acknowledgements The authors thank the entire Palomar Observatory staff for their tireless support of our work. We additionally acknowledge Jessie Christiansen for helpful discussions on KOI-1783, B. J. Fulton for assistance with the California Kepler Survey dataset, Erik Petigura for useful comments on time-correlated noise and joint RV-TTV modeling, Nicole Wallack for discussions on light curve fitting, and Gudmundur Stefansson for conversations regarding diffuser-assisted photometry at Palomar and other observatories. Support for this program was provided by NSF Career grant 1555095 and by NASA Origins grant NNX14AD22G. This work was partially supported by funding from the Center for Exoplanets and Habitable Worlds, which is supported by the Pennsylvania State University, the Eberly College of Science, and the Pennsylvania Space Grant Consortium. This research has made use of the NASA Exoplanet Archive, which is operated by the California Institute of Technology, under contract with the National Aeronautics and Space Administration under the Exoplanet Exploration Program.

\facilities{Hale (WIRC, PHARO), Kepler, OHP:1.52m (SOPHIE), PO:1.5m (ROBO-AO), Keck:II (NIRC2), ADS, Exoplanet Archive}

\software{photutils \citep{b16b}, numpy \citep{v11}, astropy \citep{a13, a18}, scipy \citep{j01}, matplotlib \citep{h07}, batman \citep{k15b}, emcee \citep{fm13}, corner \citep{fm16}, PyKE \citep{sb12}, Aladin Lite \citep{b00, b14}}

\appendix 
\section{\textit{Kepler} and WIRC Light Curves \label{ap:lightcurve}}
\begin{figure}[h!]
    \centering
    \includegraphics[width=\textwidth]{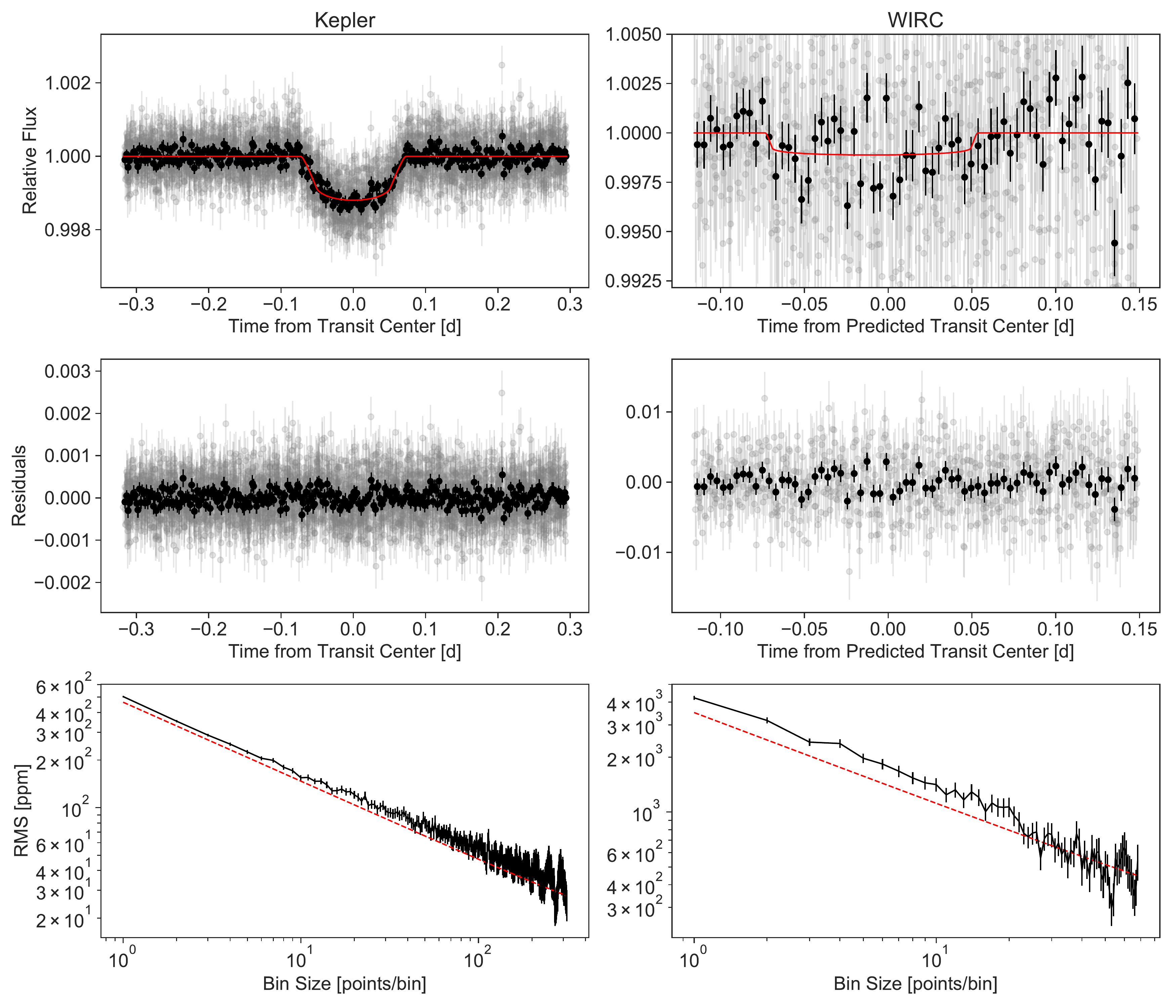}
    \caption{\textit{Kepler} (left) and WIRC (right) light curves and best-fit models (top), residuals (middle), and RMS as a function of bin size (bottom) for Kepler-29b. In the top and middle plots, the unbinned data are shown as gray filled circles, and the light curves binned by a factor of 10 are shown as black filled circles. The red lines in the top plots denote our best-fit light curve model. The transit is detected at 3.5$\sigma$ confidence in the WIRC data, and we constrain the transit timing offset to be -14$^{+17}_{-3}$ minutes (from the predicted time in Table~\ref{table1}). For continuous data acquisition with WIRC, a bin size of 24 points is equivalent to 10 minutes in the lower right plot.}
    \label{kep29fit}
\end{figure}

\begin{figure}
    \centering
    \includegraphics[width=\textwidth]{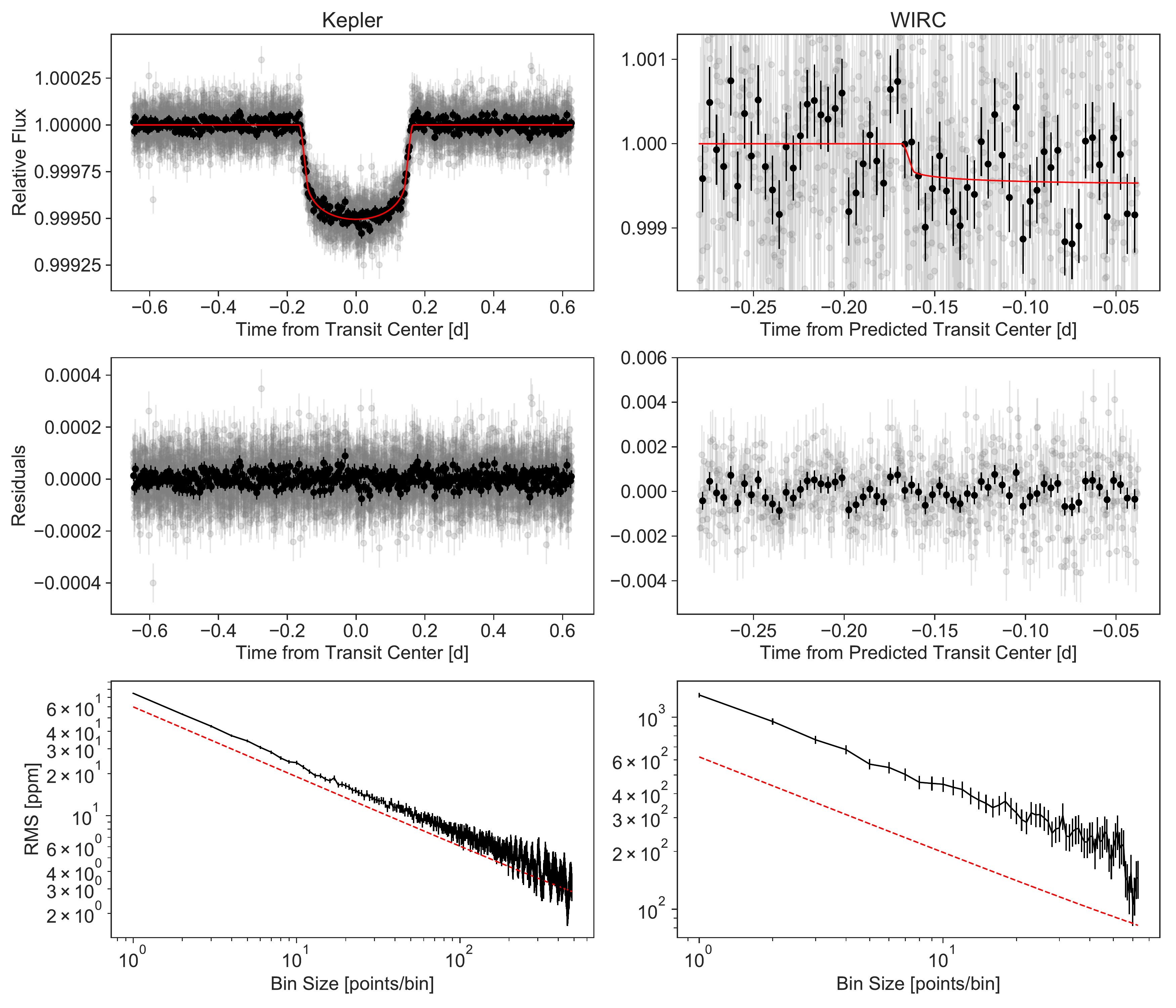}
    \caption{Same as Figure~\ref{kep29fit}, but for Kepler-36c. The transit is detected at 5.3$\sigma$ confidence in the WIRC data, and we constrain the transit timing offset to be -18$^{+12}_{-5}$ minutes (from the predicted time in Table~\ref{table1}). For continuous data acquisition with WIRC, a bin size of 38 points is approximately equivalent to 10 minutes in the lower right plot.}
    \label{kep36fit}
\end{figure}

\begin{figure}
    \centering
    \includegraphics[width=\textwidth]{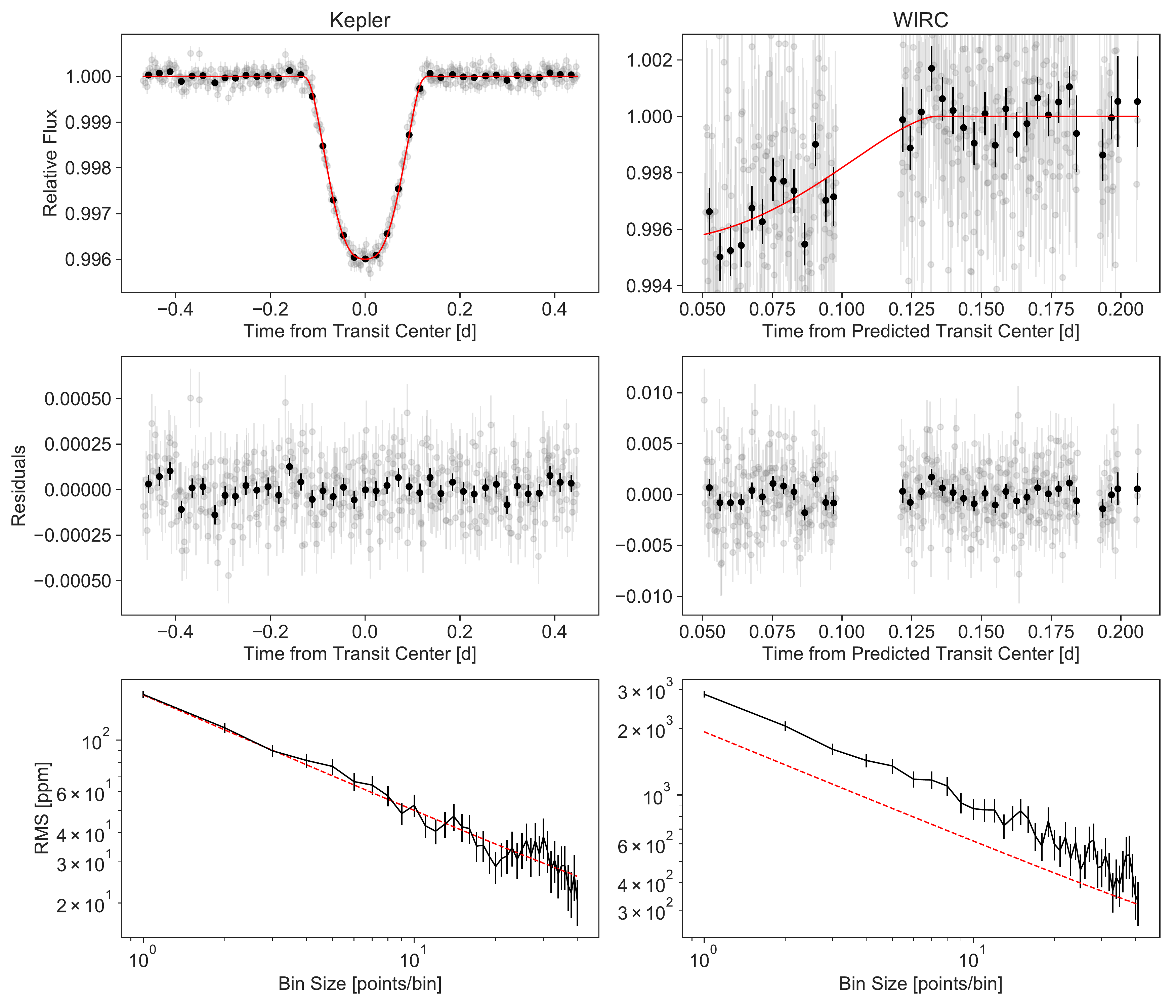}
    \caption{Same as Figure~\ref{kep29fit}, but for KOI-1783.01. The transit is detected at 5.9$\sigma$ confidence in the WIRC data, and we constrain the transit timing offset to be 16$^{+10}_{-11}$ minutes (from the predicted time in Table~\ref{table1}). For continuous data acquisition with WIRC, a bin size of 30 points is equivalent to 10 minutes in the lower right plot (note however the breaks in data acquisition).}
    \label{koi1783fit}
    
\end{figure}

\begin{figure}
    \centering
    \includegraphics[width=\textwidth]{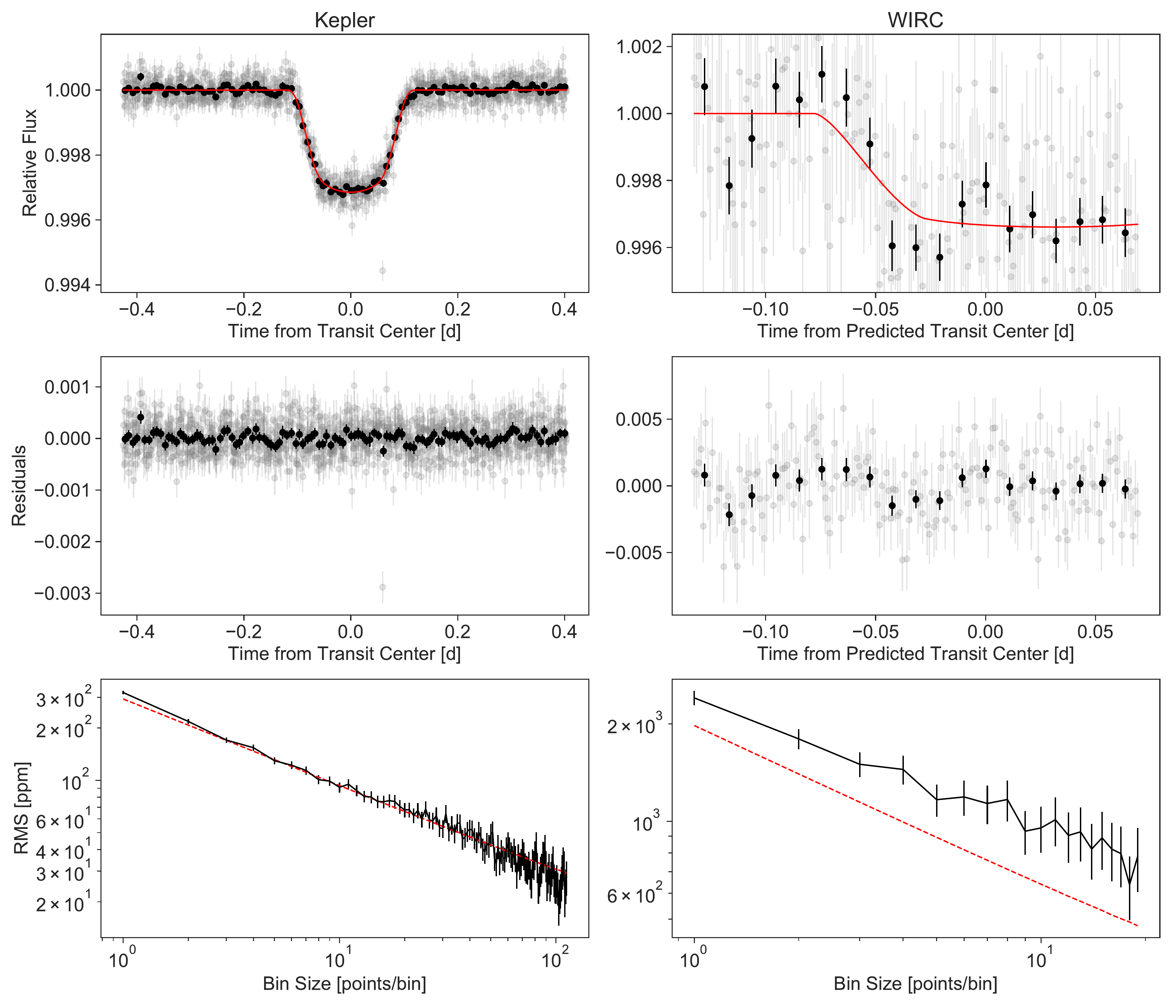}
    \caption{Same as Figure~\ref{kep29fit}, but for Kepler-177c. The transit is detected at 5.5$\sigma$ confidence in the WIRC data, and we constrain the transit timing offset to be 45$^{+9}_{-7}$ minutes (from the predicted time in Table~\ref{table1}). For continuous data acquisition with WIRC, a bin size of 8 points is equivalent to 10 minutes in the lower right plot (note however the breaks in data acquisition in this observation).}
    \label{kep177fit}
\end{figure}

\clearpage
\section{Posterior Probability Distributions \label{ap:posteriors}}
\begin{figure}[h!]
    \centering
    \includegraphics[width=0.9\textwidth]{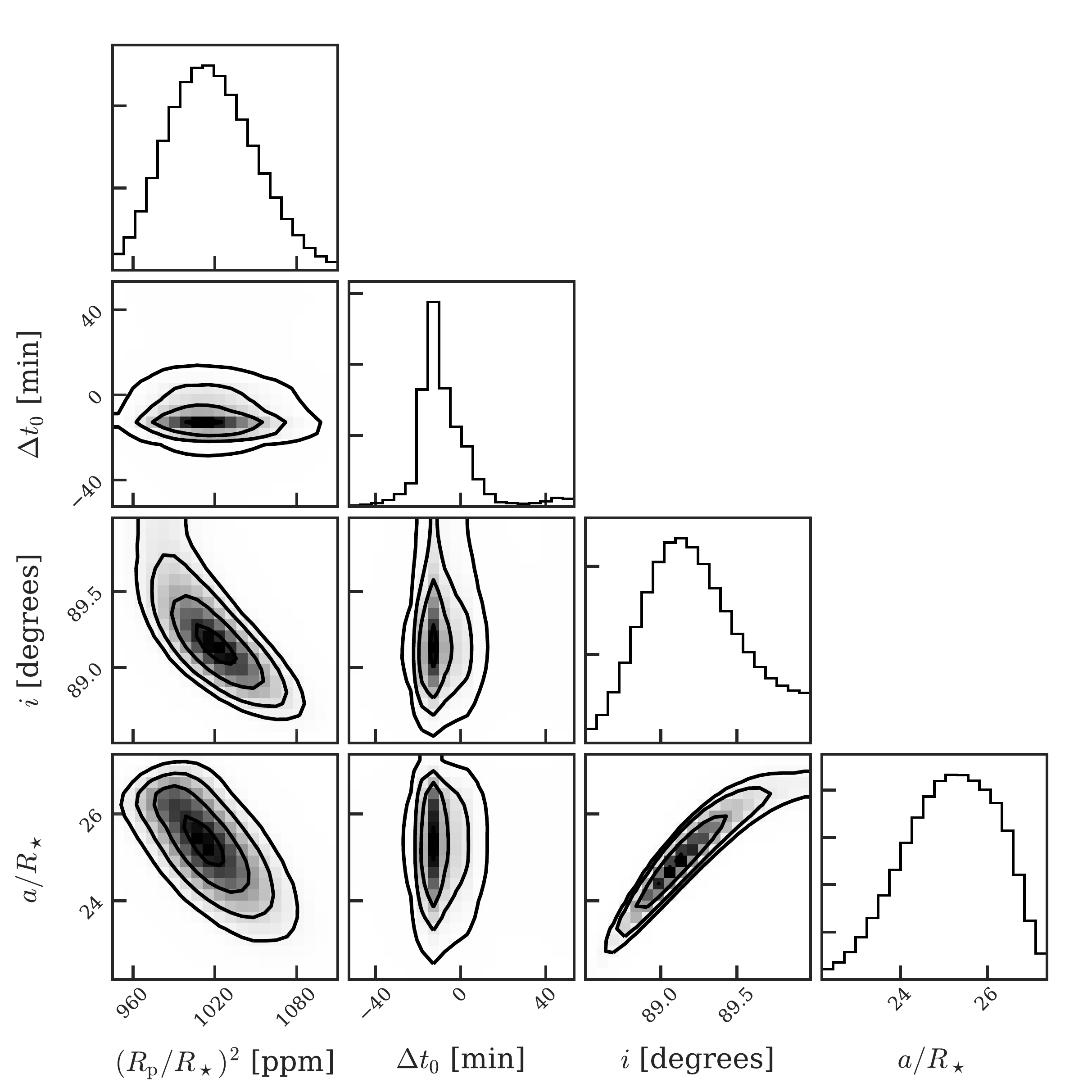}
    \caption{The posterior probability distributions for our fit to Kepler-29b. For ease of viewing, only the middle 99 percent of the samples are shown for each distribution, and the contours denote 1, 2, and 3$\sigma$ boundaries.}
    \label{kep29corner}
\end{figure}

\begin{figure}
    \centering
    \includegraphics[width=0.9\textwidth]{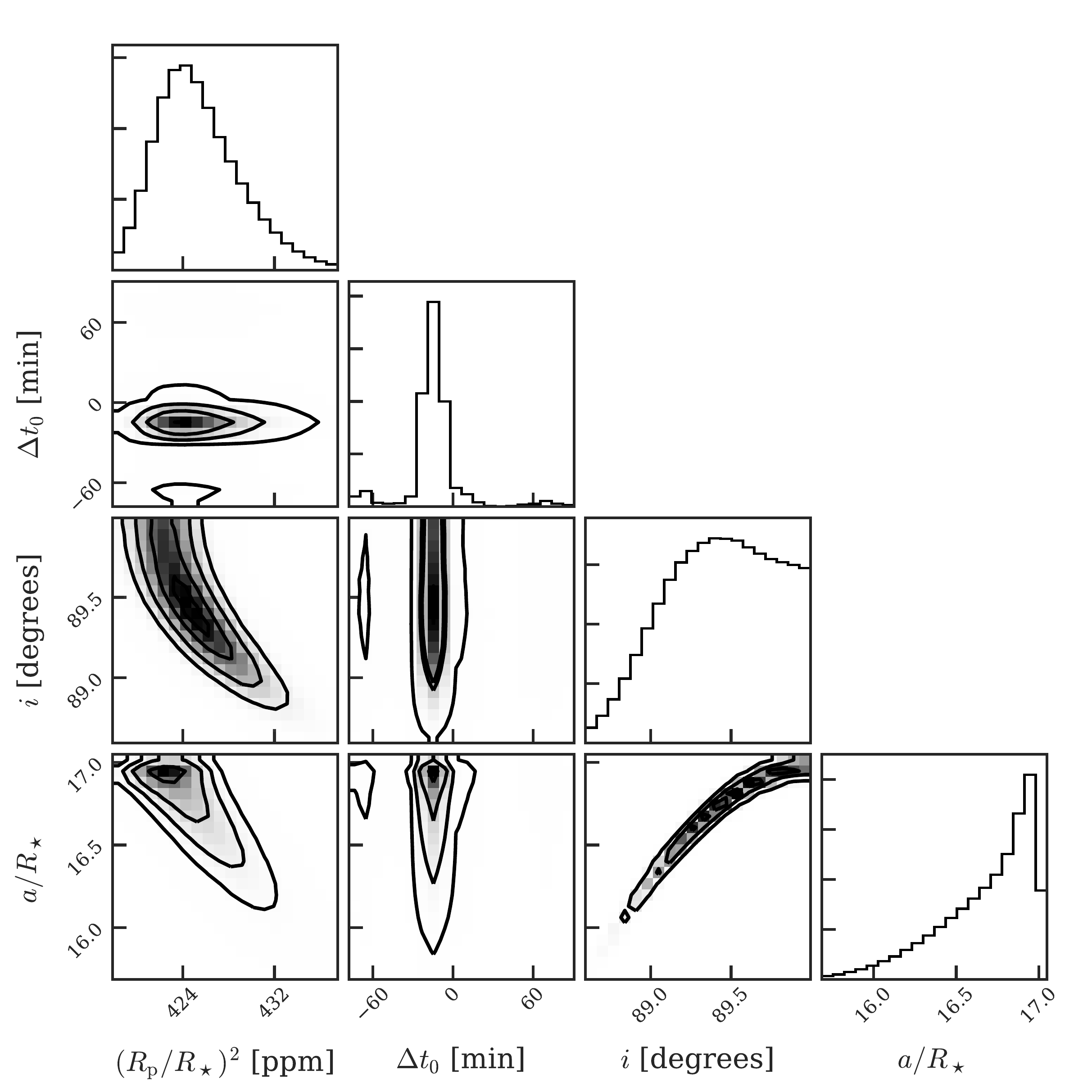}
    \caption{Same as Figure~\ref{kep29corner}, but for Kepler-36c.}
    \label{kep36corner}
\end{figure}

\begin{figure}
    \centering
    \includegraphics[width=0.9\textwidth]{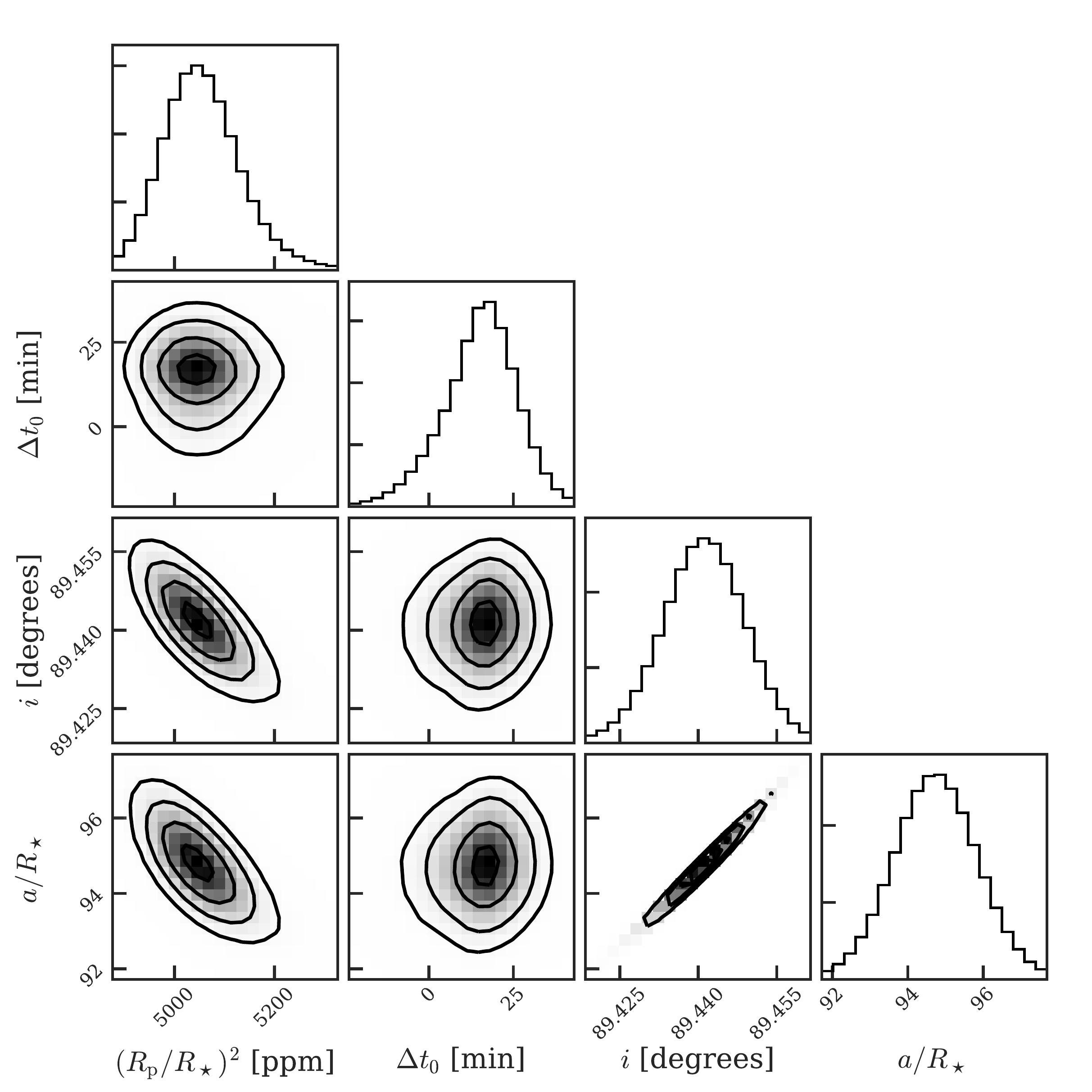}
    \caption{Same as Figure~\ref{kep29corner}, but for KOI-1783.01.}
    \label{koi1783corner}
\end{figure}

\begin{figure}
    \centering
    \includegraphics[width=0.9\textwidth]{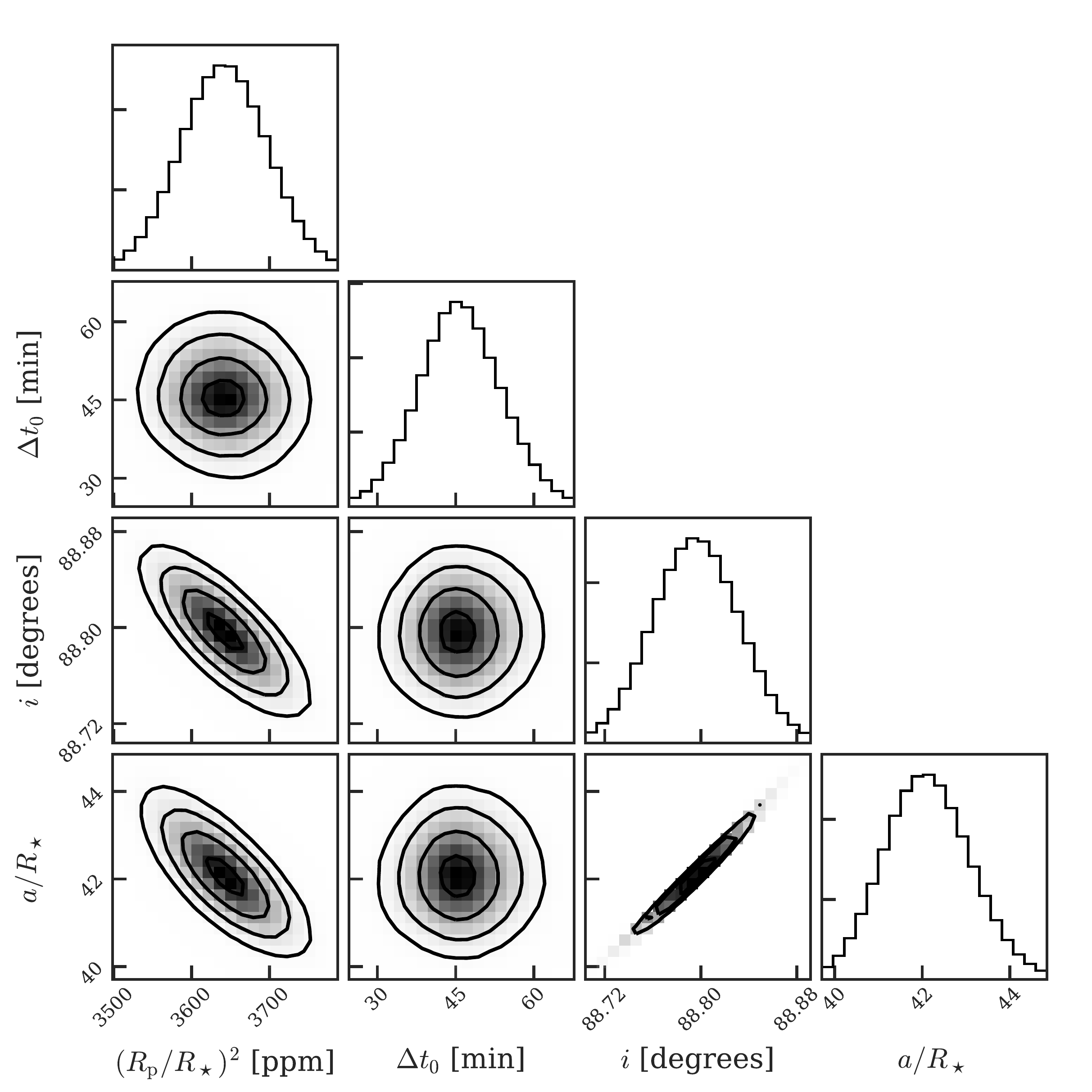}
    \caption{Same as Figure~\ref{kep29corner}, but for Kepler-177c.}
    \label{kep177corner}
\end{figure}

\clearpage
\section{Dynamical Modeling Results \label{ap:dynamicalmodeling}}

\begin{figure}[h!]
    \centering
    \gridline{\fig{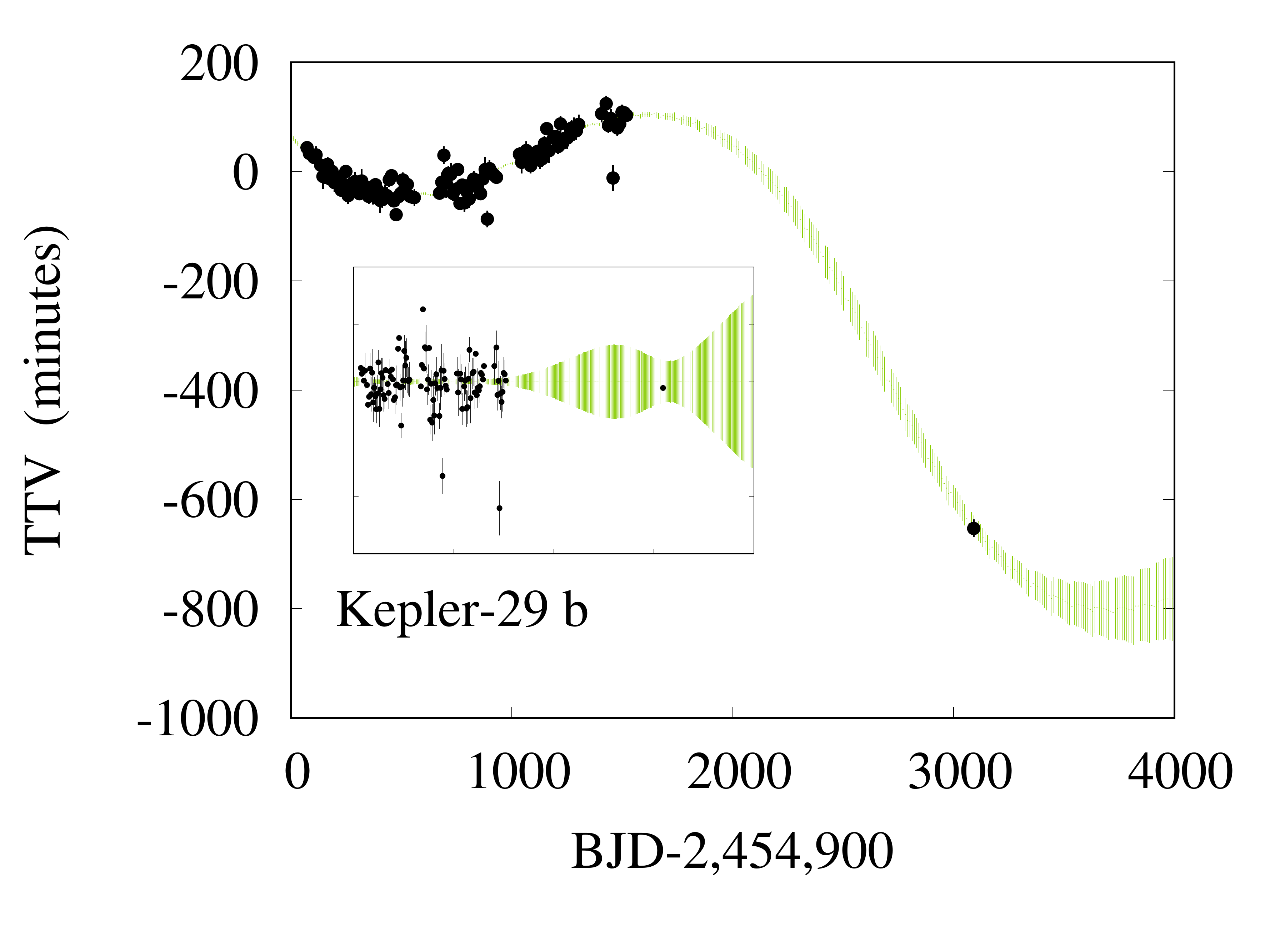}{0.47\textwidth}{(a)}
          \fig{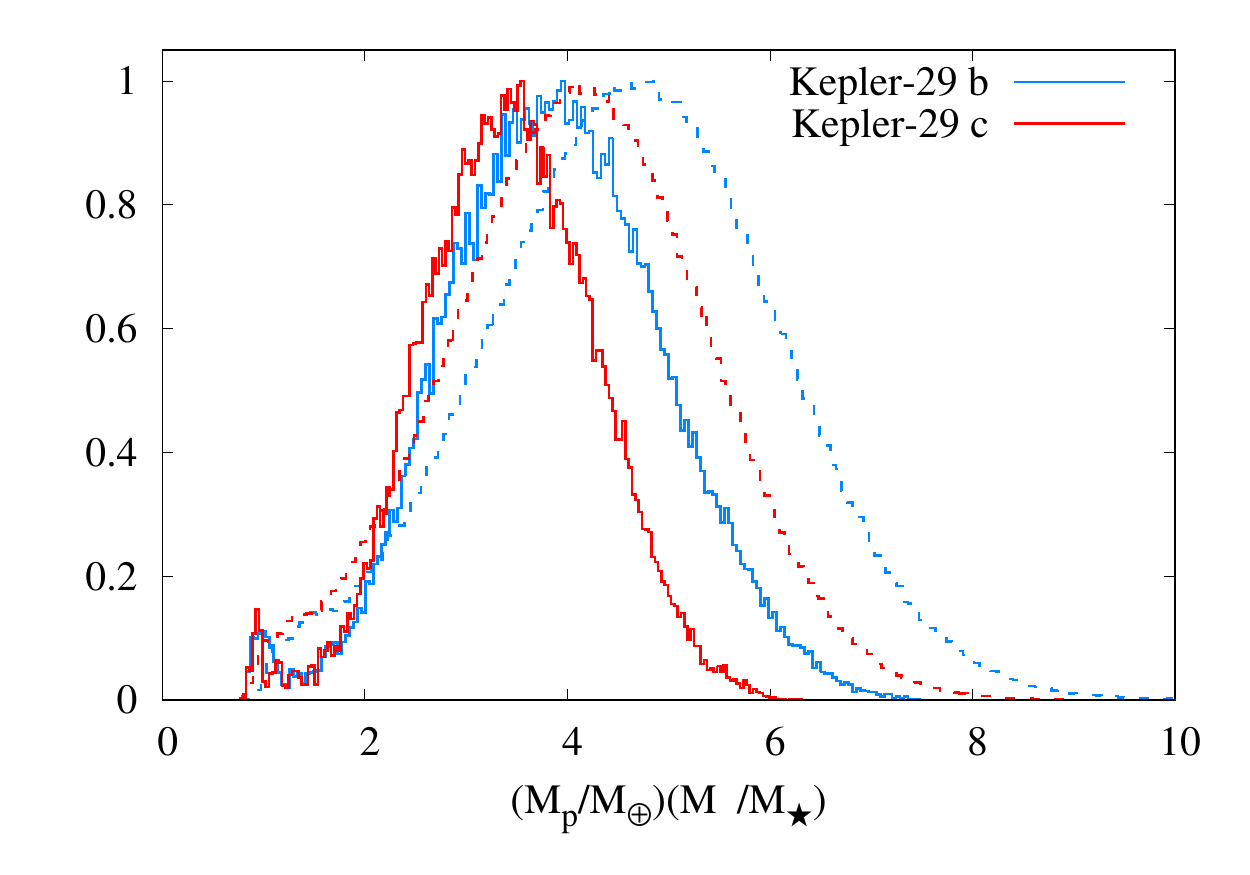}{0.47\textwidth}{(b)}}
    \gridline{\fig{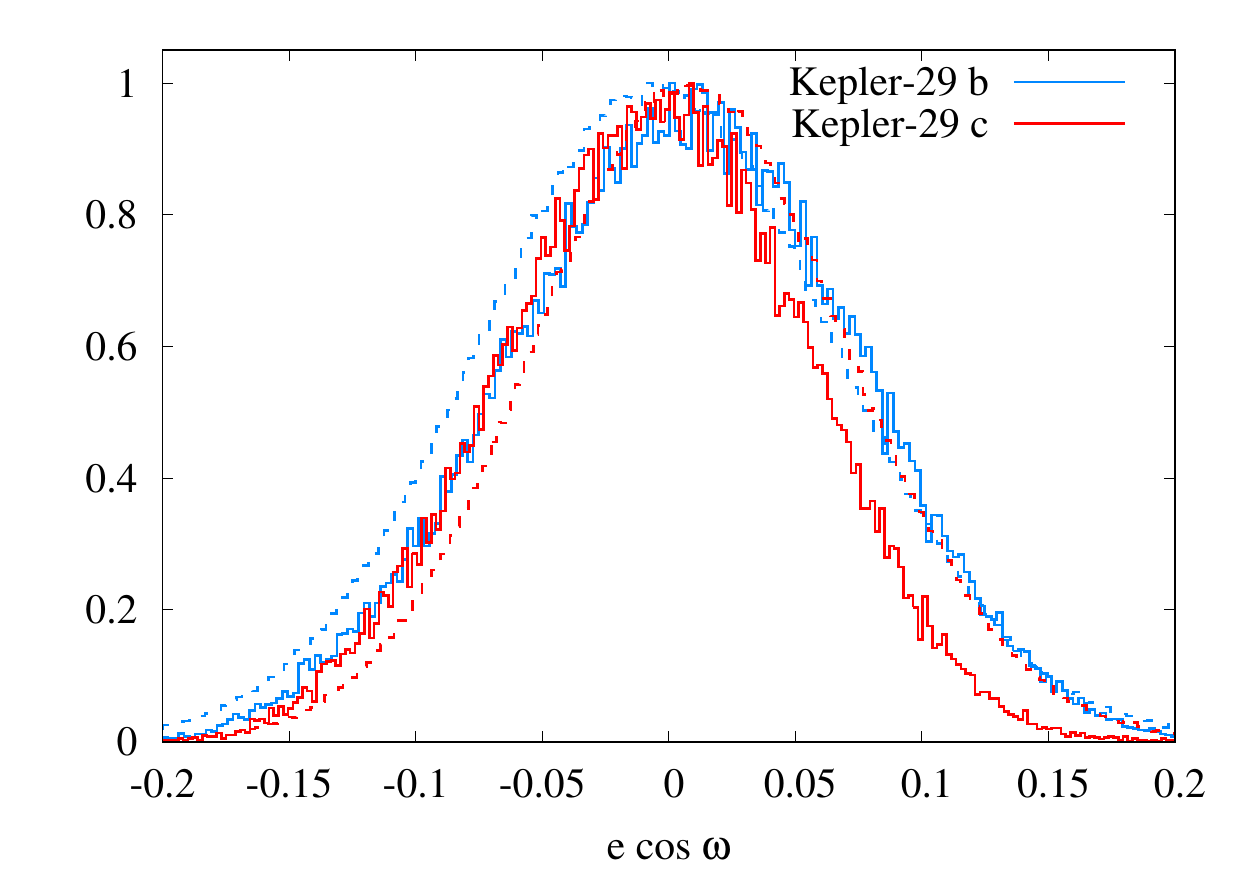}{0.47\textwidth}{(c)}
          \fig{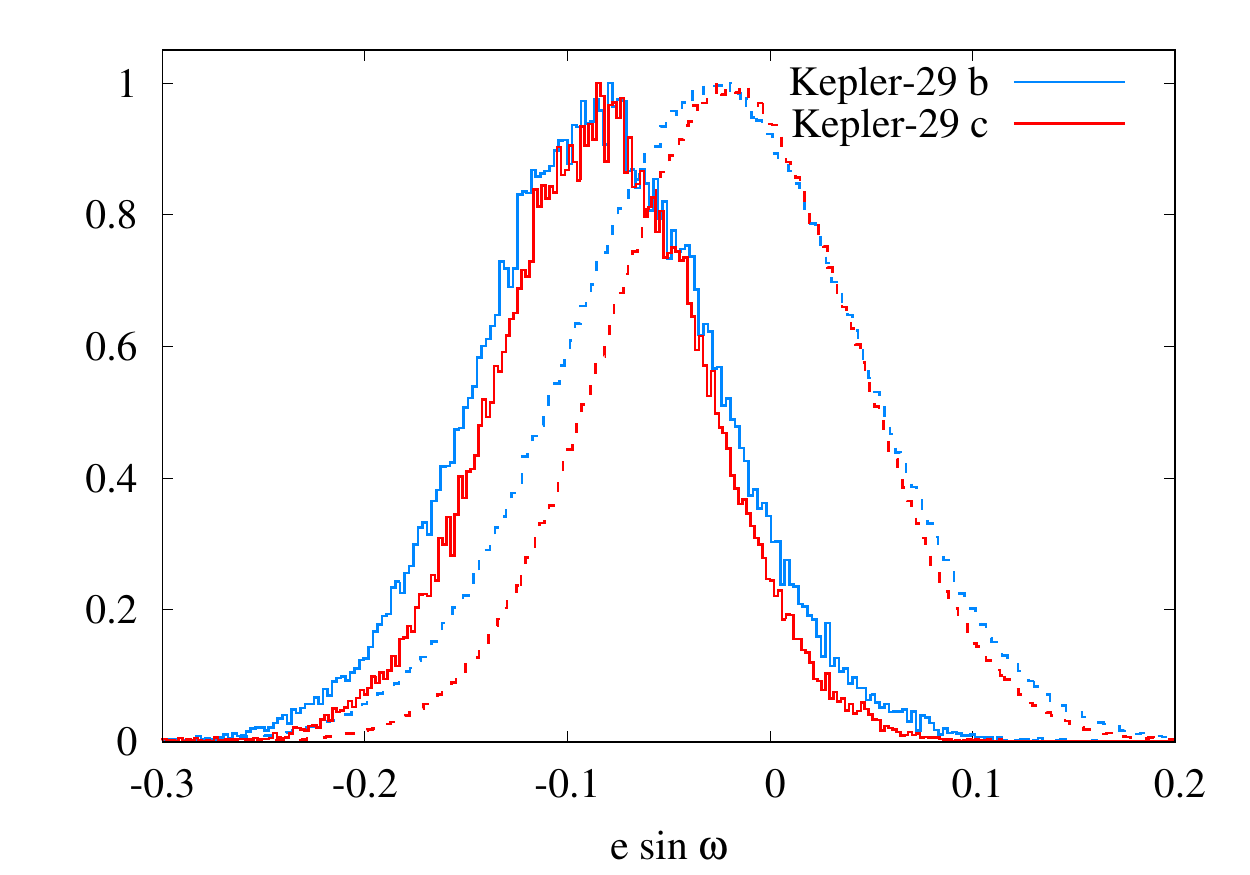}{0.47\textwidth}{(d)}}
    \caption{Updated dynamical modeling of the Kepler-29 system based on fits to \textit{Kepler} and WIRC transit times. (a) The measured transit timing variations (i.e., deviations from a constant ephemeris using the period derived from our TTV modeling) for Kepler-29b from the \textit{Kepler} and WIRC transit observations (black filled circles); we also overplot the 1$\sigma$ range in predicted TTVs for each epoch from the updated dynamical model in green. We include an inset of the residuals from the best fit TTV model to show how our new measurement compares to the \textit{Kepler} uncertainties. (b) The dynamical mass posteriors for both planets in the system. (c and d) The posteriors on both components of the eccentricity vectors. Posteriors from TTV modeling of the \textit{Kepler} data are shown as dashed lines, and those from joint modeling of the \textit{Kepler} and WIRC data are shown as solid lines.}
    \label{kep29data}
\end{figure}

\begin{figure}
    \centering
    \gridline{\fig{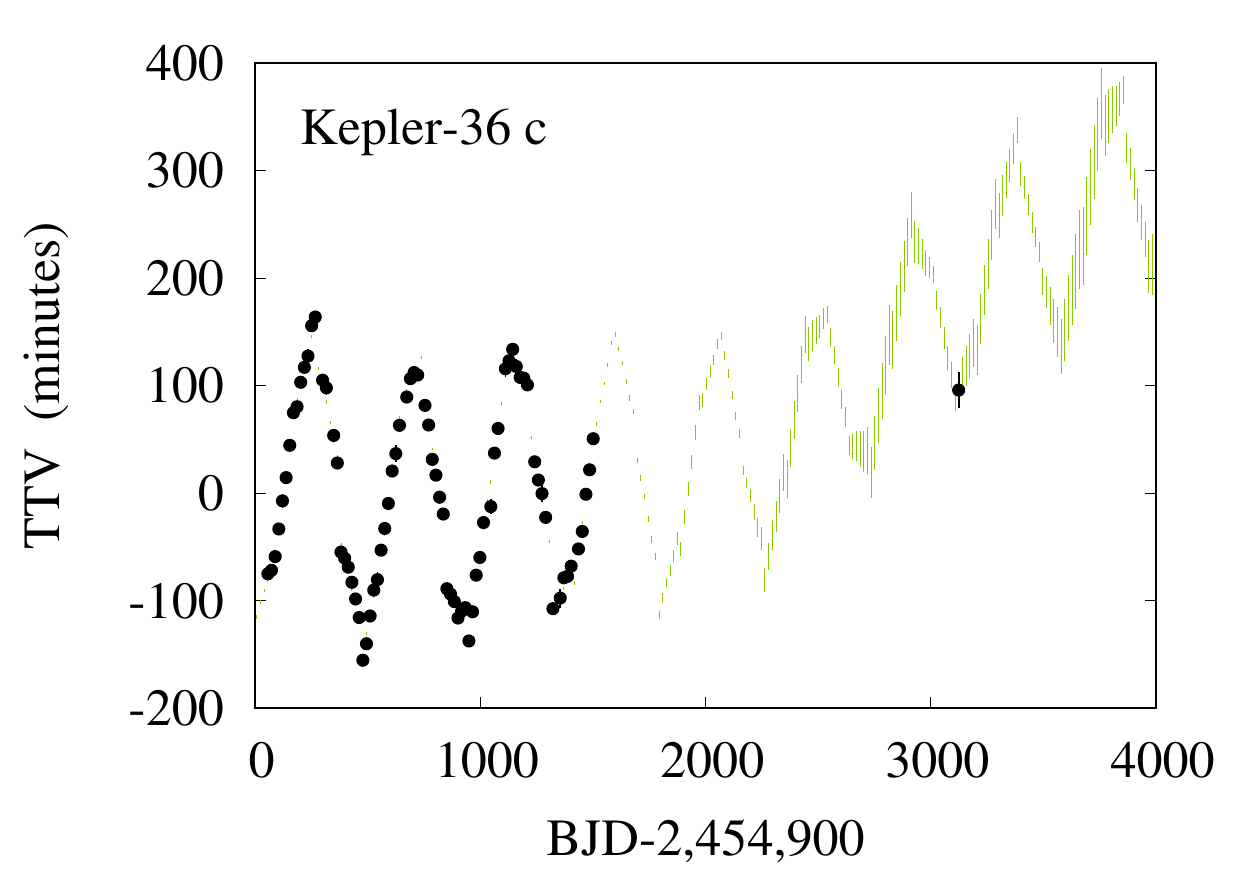}{0.47\textwidth}{(a)}
          \fig{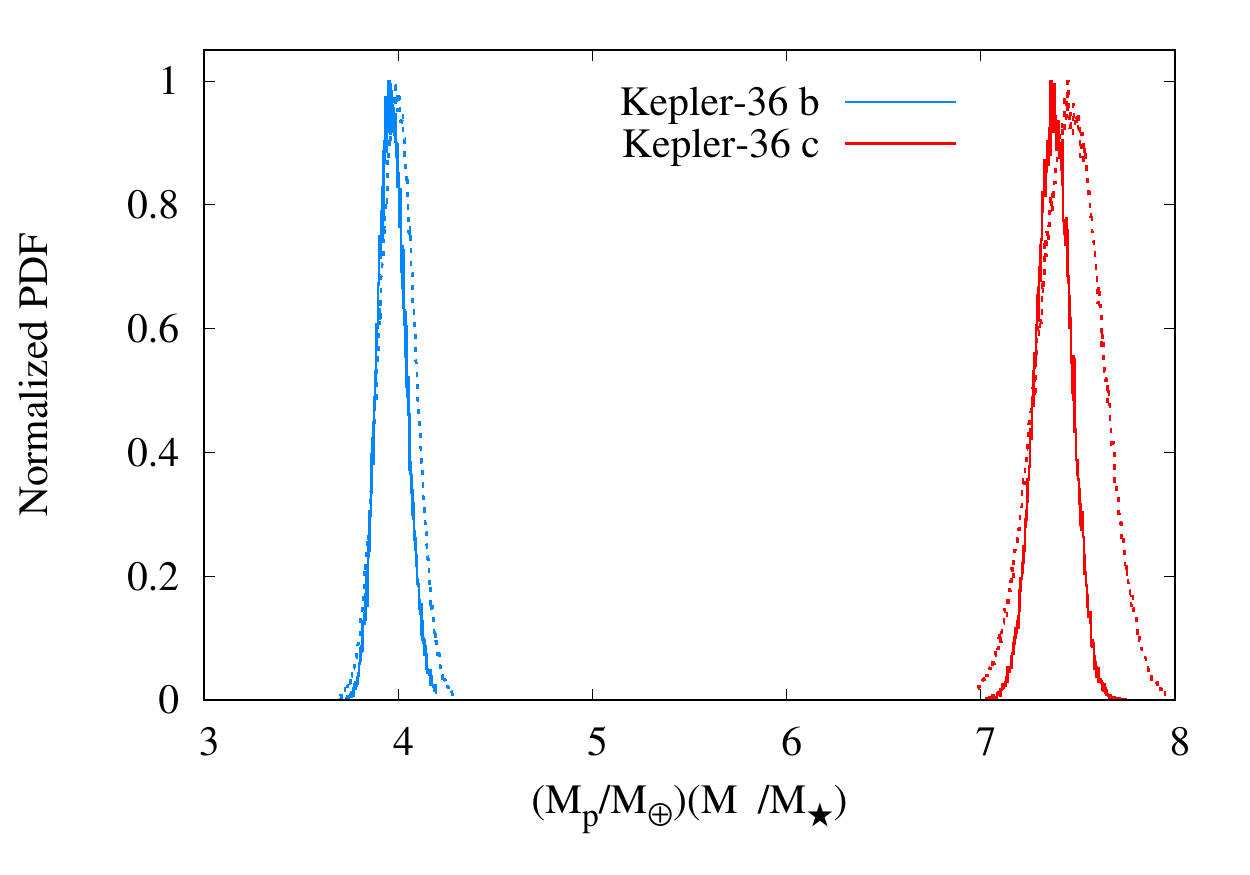}{0.47\textwidth}{(b)}}
    \gridline{\fig{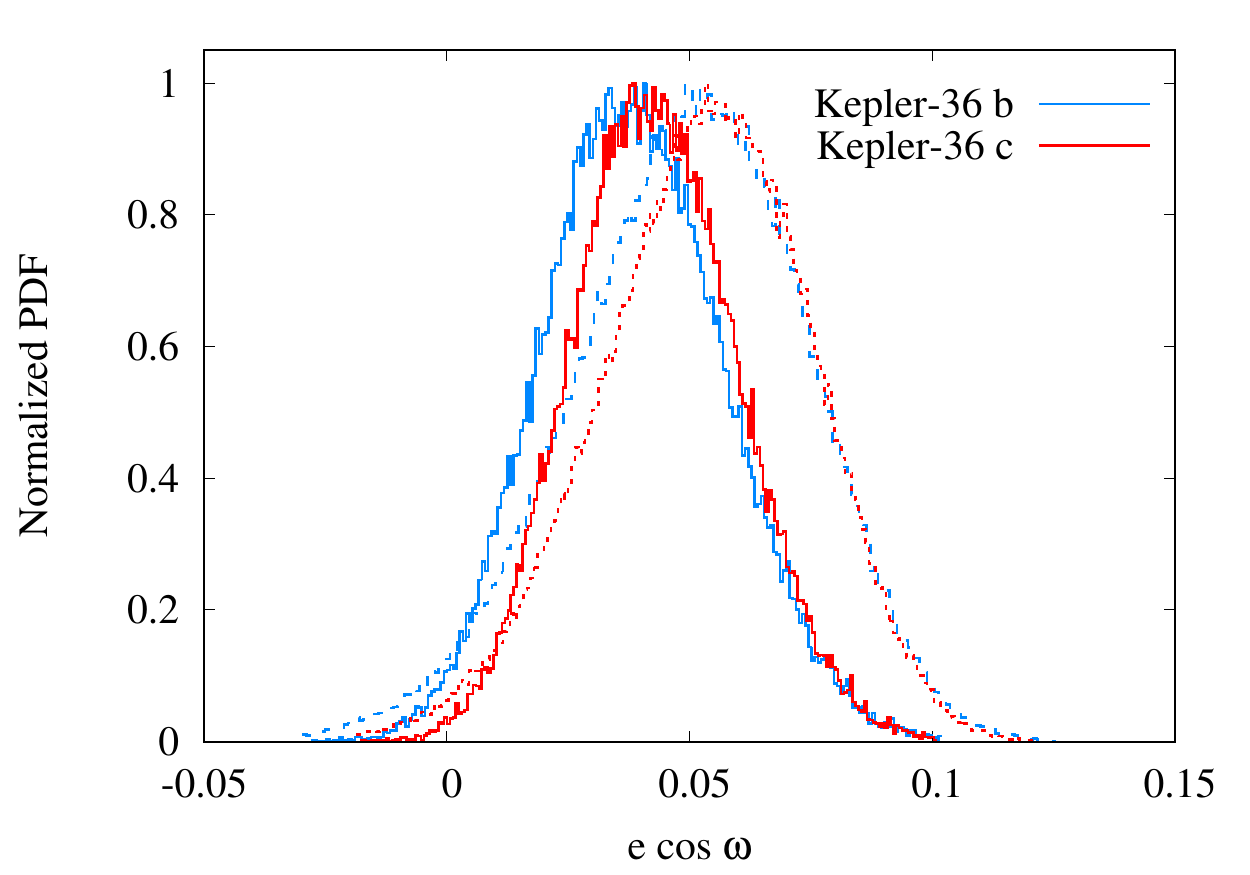}{0.47\textwidth}{(c)}
          \fig{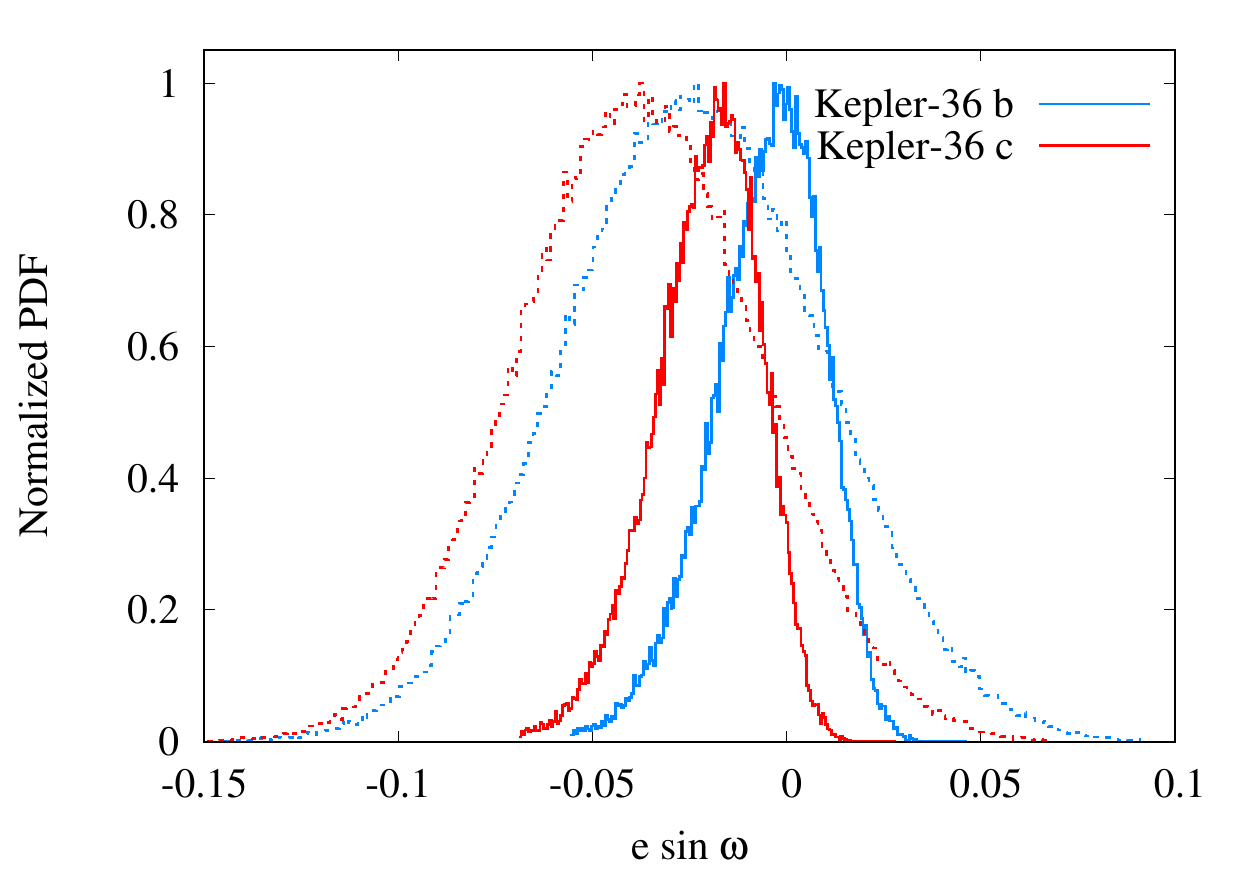}{0.47\textwidth}{(d)}}
    \caption{Same as Figure~\ref{kep29data}, but for Kepler-36.}
    \label{kep36data}
\end{figure}

\begin{figure}
    \centering
    \gridline{\fig{{koi1783.01}.pdf}{0.47\textwidth}{(a)}
          \fig{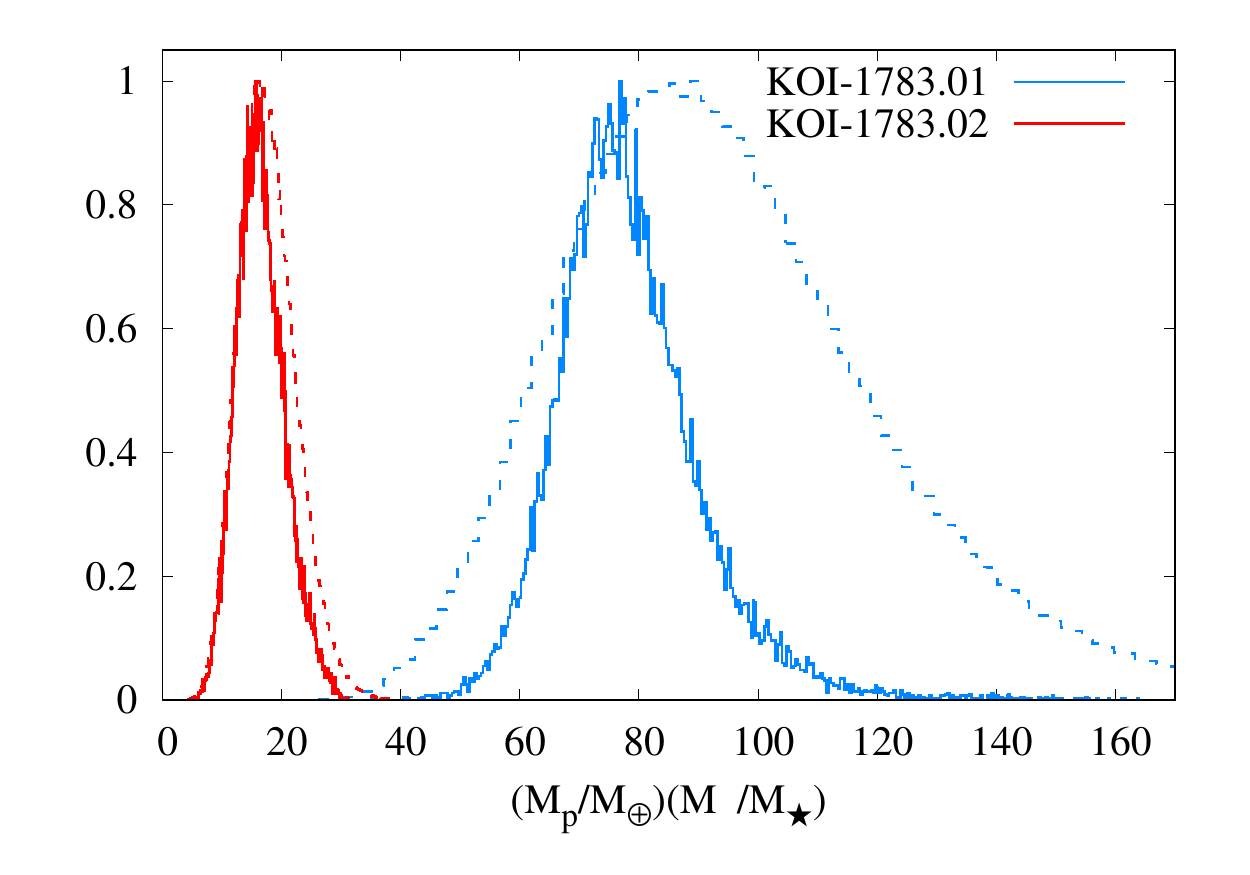}{0.47\textwidth}{(b)}}
    \gridline{\fig{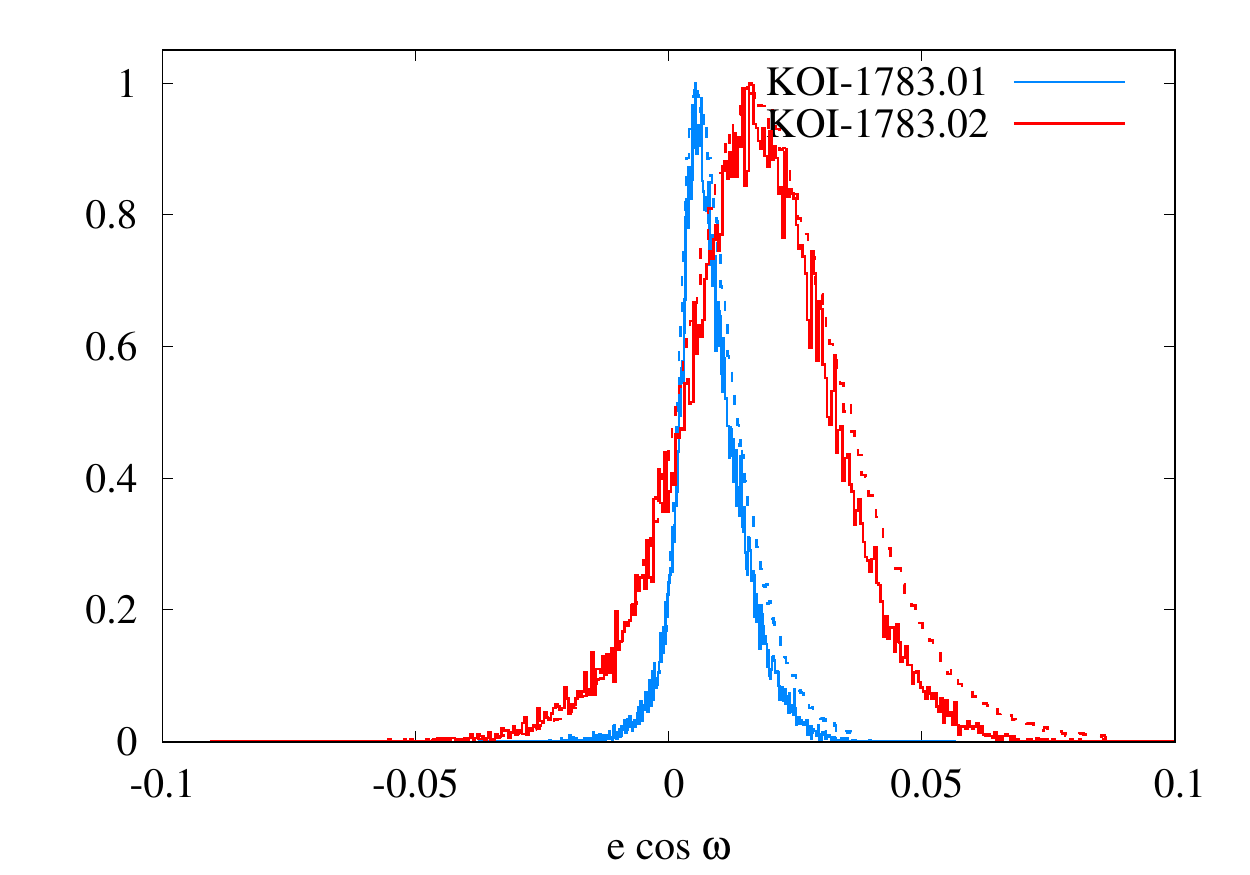}{0.47\textwidth}{(c)}
          \fig{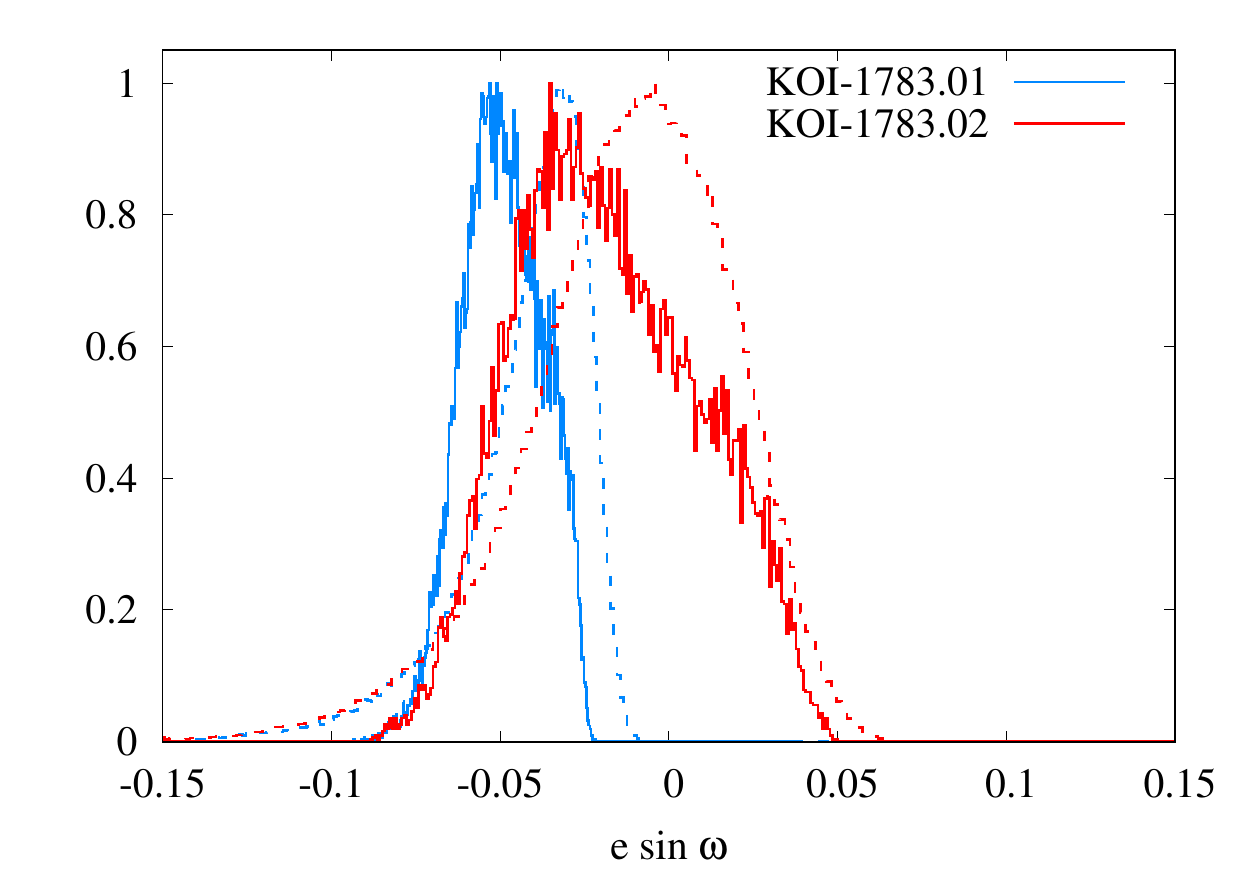}{0.47\textwidth}{(d)}}
    \caption{Same as Figure~\ref{kep29data}, but for KOI-1783.}
    \label{koi1783data}
\end{figure}

\begin{figure}
    \centering
    \gridline{\fig{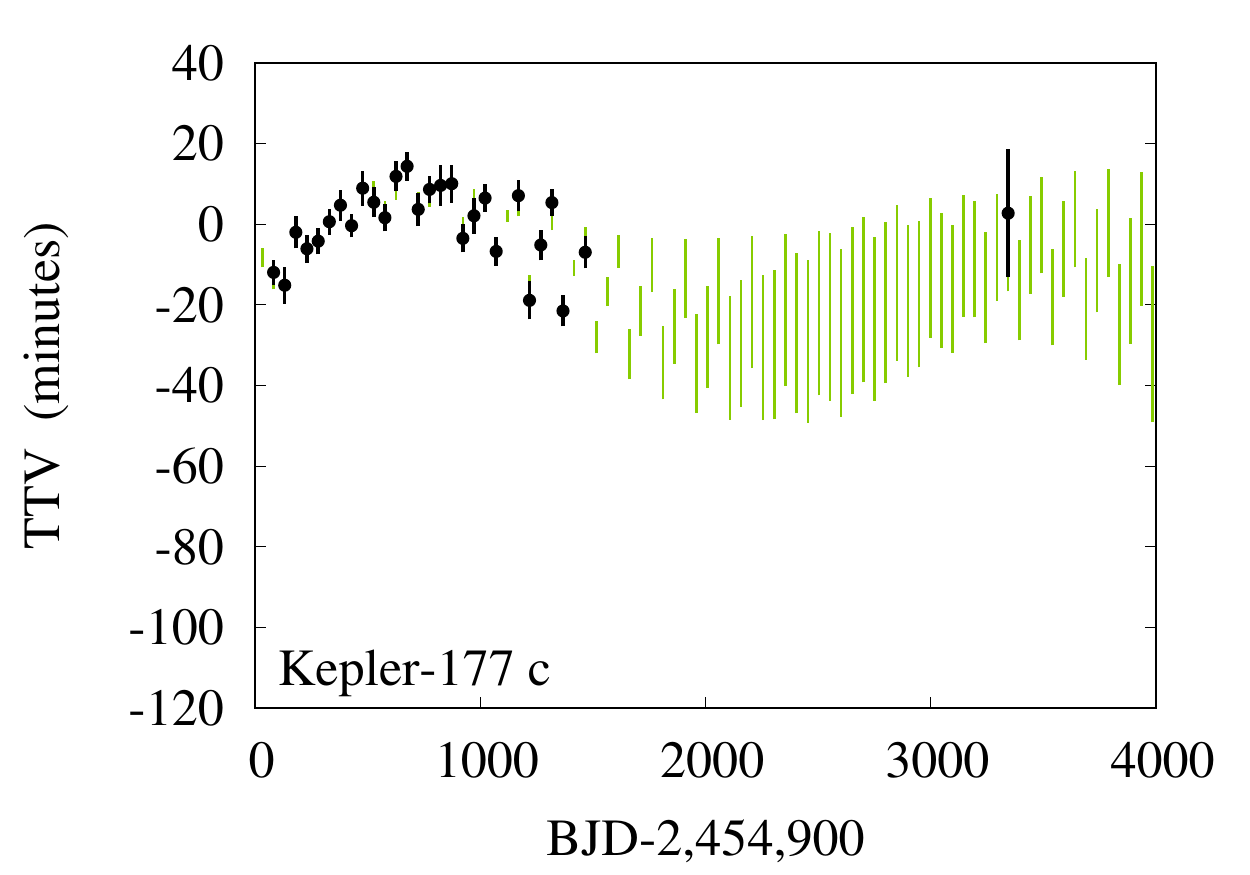}{0.47\textwidth}{(a)}
          \fig{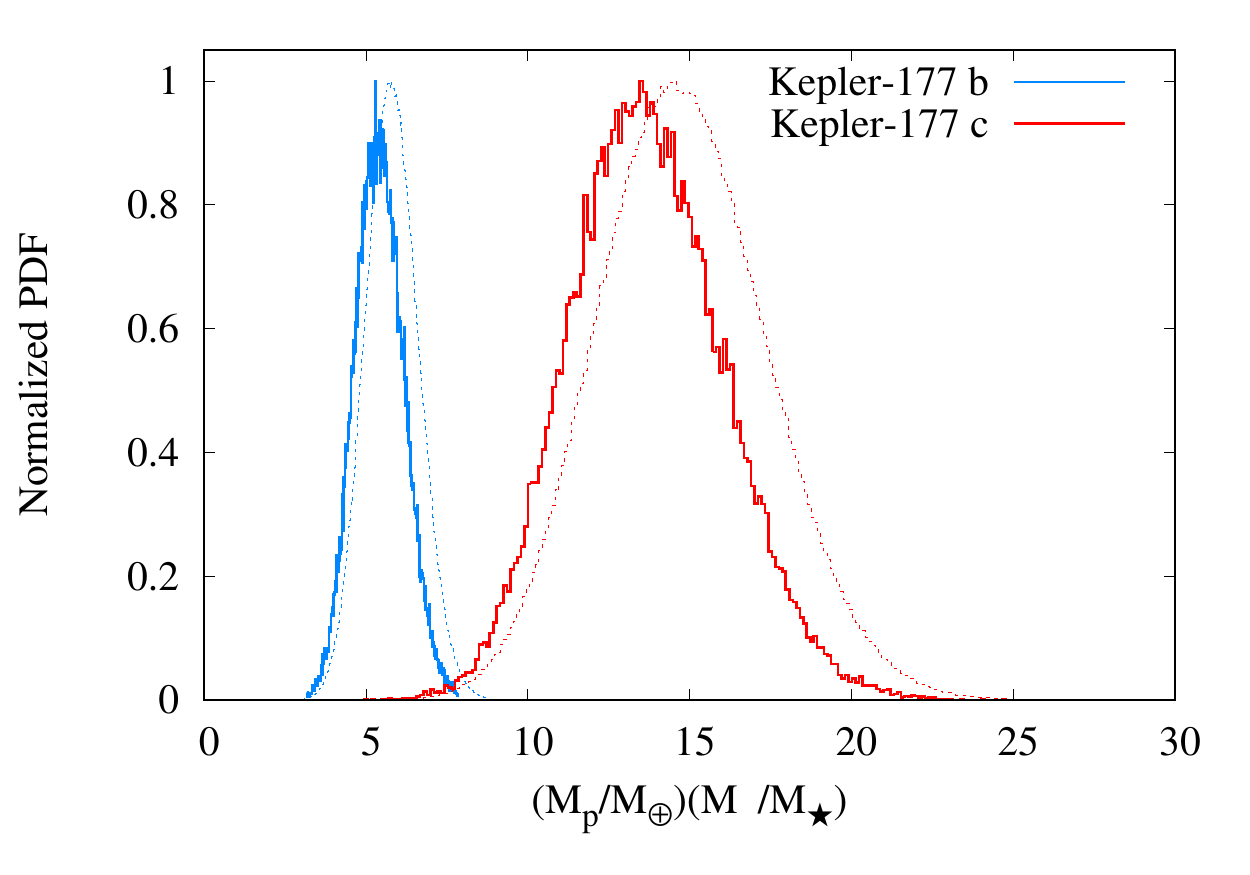}{0.47\textwidth}{(b)}}
    \gridline{\fig{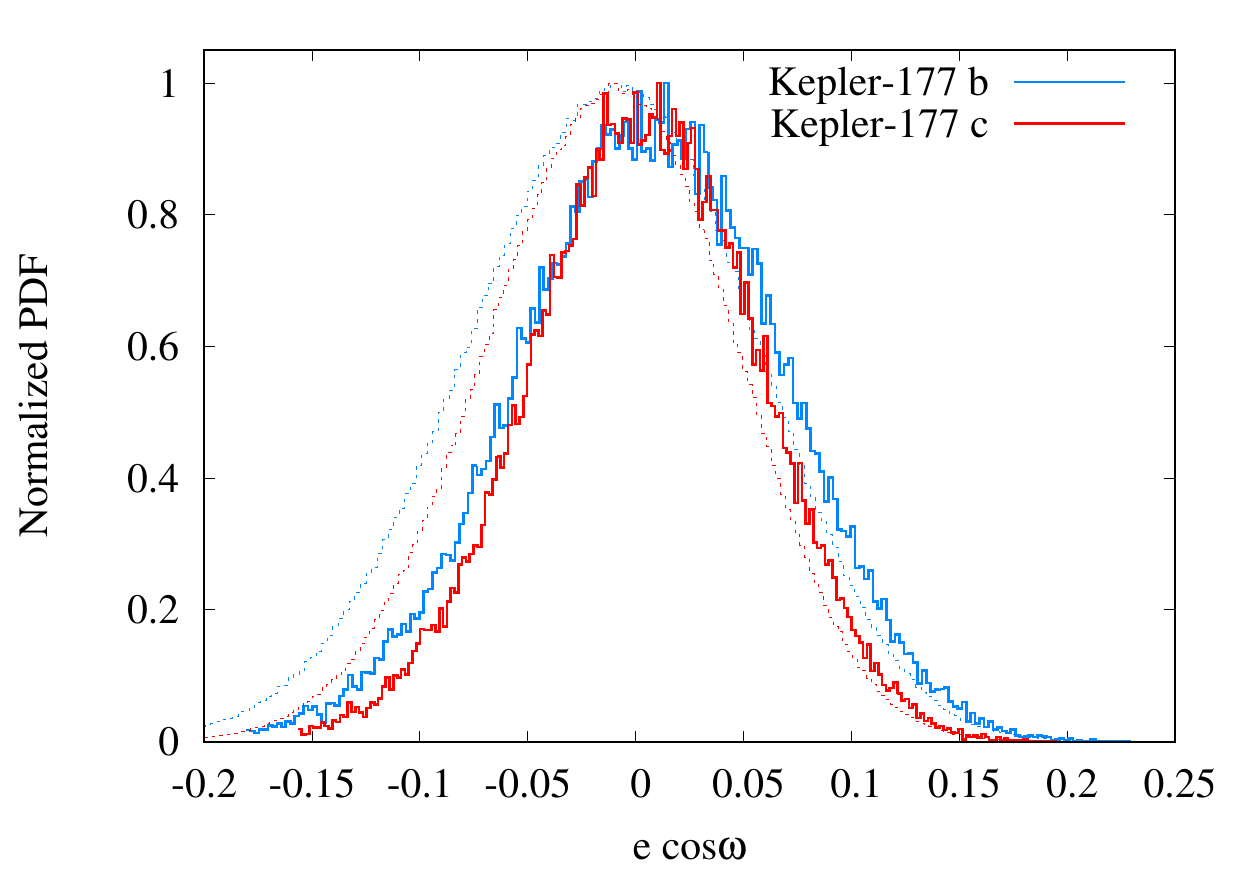}{0.47\textwidth}{(c)}
          \fig{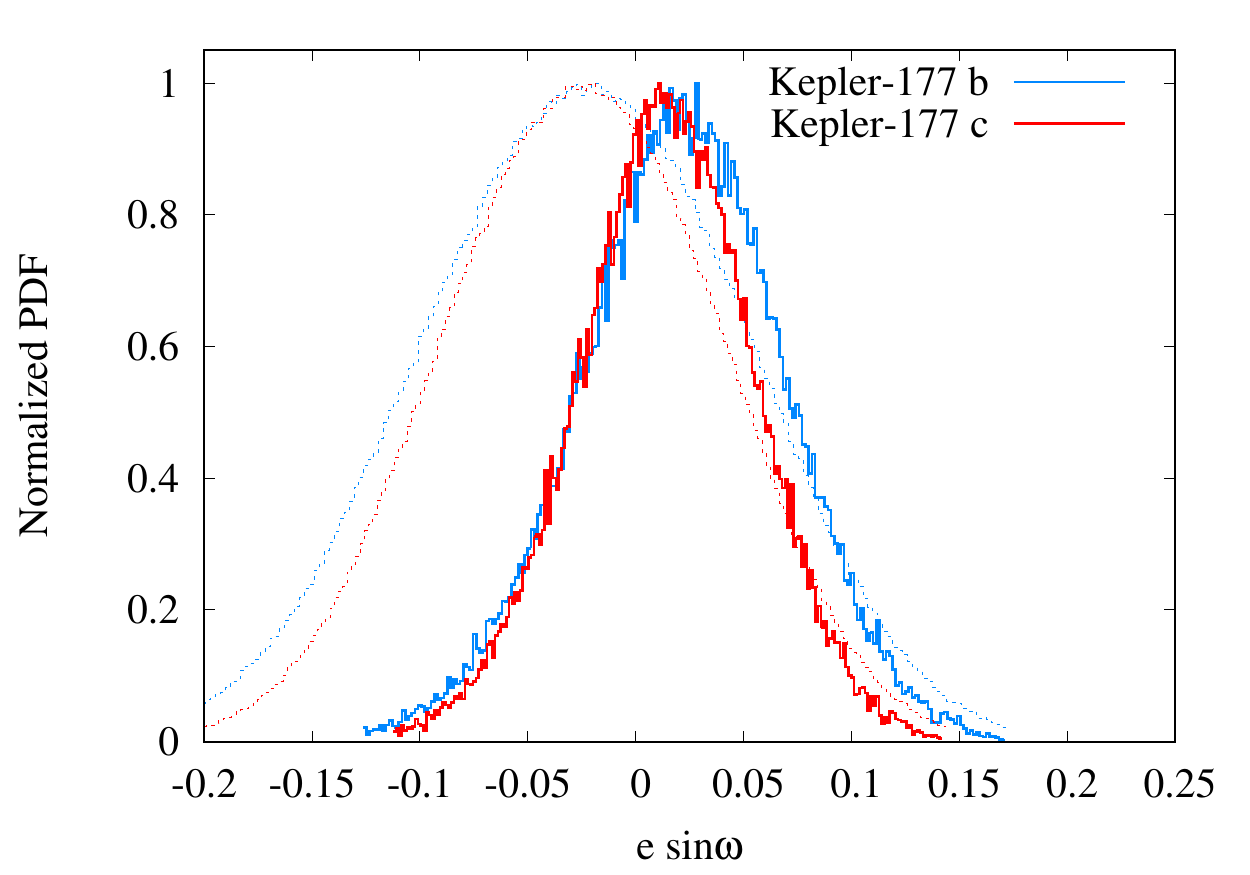}{0.47\textwidth}{(d)}}
    \caption{Same as Figure~\ref{kep29data}, but for Kepler-177.}
    \label{kep177data}
\end{figure}

\end{document}